\documentclass[
preprint,
 amsmath,amssymb,
floatfix,
]{revtex4-1}

\usepackage{graphicx}
\usepackage{dcolumn}
\usepackage{bm}
\usepackage[separate-uncertainty=true]{siunitx}
\RequirePackage{lineno}
\DeclareSIUnit{\molar}{M}
\usepackage{tabularx}
\usepackage{longtable}
\usepackage{booktabs}
\usepackage{physics}
\usepackage[version=4]{mhchem}
\usepackage{float}
\usepackage{bold-extra}
\usepackage{xcolor, soul} 
\usepackage[export]{adjustbox}
\usepackage{makecell}

\usepackage[colorinlistoftodos]{todonotes}
\DeclareSIUnit{\Molar}{\textsc{m}}
\DeclareSIUnit{\bp}{bp}
\DeclareSIUnit{\nt}{nt}
\DeclareSIUnit{\pH}{pH}
\DeclareSIUnit{\calorie}{cal}
\DeclareSIUnit{\mole}{mol}
\DeclareSIUnit{\min}{min}
\DeclareSIUnit{\kT}{k_BT}
\DeclareSIUnit\angstrom{\protect \text {Å}}
\begin{document}

\title{Weak tension accelerates hybridization and dehybridization of short oligonucleotides}
\author{
Derek J. Hart, Jiyoun Jeong, James C. Gumbart, Harold D. Kim\footnote{To whom correspondence should be addressed. Tel: +1 404 8940080; Fax: +1 404 8941101; Email: harold.kim@physics.gatech.edu}}
\address{School of Physics, Georgia Institute of Technology, 837 State Street, Atlanta, GA 30332-0430, USA}

\date{\today}
\begin{abstract}

The hybridization and dehybridization of DNA subject to tension is relevant to fundamental genetic processes and to the design of DNA-based mechanobiology assays. While strong tension accelerates DNA melting and decelerates DNA annealing, the effects of tension weaker than \SI{5}{\pico\newton} are less clear. In this study, we developed a DNA bow assay, which uses the bending rigidity of double-stranded DNA (dsDNA) to exert weak tension on a single-stranded DNA (ssDNA) target in the range of \SIrange{2}{6}{\pico\newton}.  Combining this assay with single-molecule FRET, we measured the hybridization and dehybridization kinetics between a \SI{15}{\nt} ssDNA under tension  and a 8-9 \SI{}{\nt} oligo, and found that both the hybridization and dehybridization rates monotonically increase with tension for various nucleotide sequences tested. These findings suggest that the nucleated duplex in its transition state is more extended than the pure dsDNA or ssDNA counterpart. Our simulations using the coarse-grained oxDNA2 model indicate that the increased extension of the transition state is due to exclusion interactions between unpaired ssDNA regions in close proximity to one another. This study highlights an example where the ideal worm-like chain models fail to explain the kinetic behavior of DNA in the low force regime. 
\end{abstract}

\maketitle

\section{Introduction}
DNA strand separation or unzipping followed by annealing or rezipping is commonplace in many fundamental genomic processes such as homologous recombination and R-loop formation \cite{marmur1960strand, gai2010origin, donmez2006mechanisms, belotserkovskii2018r, aguilera2012r, alberts1994}. Although genomic processes inside the cell are orchestrated by motor proteins or enzymes, they are thought to be aided by intrinsic dynamics of the underlying genomic DNA \cite{liu2007human,choi2004dna,benham1996duplex,benham1993sites, meng2014coexistence}. Therefore, thermally-induced separation of duplex DNA into single strands and its reverse reaction may play an important role in active genomic processes. For example, in both prokaryotic and eukaryotic genomes, origins of replication commonly feature a \SIrange{10}{100}{\bp} DNA unwinding element, whose weak duplex stability determines origin function \cite{kowalski1989dna,martinez2017origin,kemp2007structure}. In CRISPR-Cas systems, melting is a rate-limiting step for Cas9 target selection, and has also been found to induce off-target binding and cleavage \cite{newton2019dna, klein2018hybridization, gong2018dna}.

The melting probability of a duplex region depends not only on its sequence \cite{zhabinskaya2012theoretical}, but also on the local stress \cite{matek2015plectoneme, zhabinskaya2012theoretical, li2019mechanism, saha2006chromatin, clapier2017mechanisms}. The genomic DNA \textit{in vivo} is seldom in a relaxed state, but rather is subjected to various forms of stress: bending, twisting, and tension. Several DNA force spectroscopy experiments have carefully explored how melting is affected by a strong artificial tension \cite{rief1999sequence, calderon2008quantifying, clausen2000mechanical,albrecht2008molecular}, but the effect of weak tension ($<\SI{5}{\pico\newton}$), which is arguably more relevant to genomic processes \textit{in vivo} or DNA-based systems \textit{in vitro}, is less clear. Forces in this range can be exerted on a duplex region during active processes such as loop extrusion by SMC complexes \cite{marko2019dna, ganji2018real} and also by thermal fluctuations of flanking DNA segments \cite{waters2015calculation}. Molecules involved in cell mechanotransduction also regularly experience forces at this scale \cite{pan2021quantifying}. Therefore, understanding the effect of weak tension on DNA hybridization/dehybridization can elucidate the physical regulation of genomic processes, and aid our design of DNA-based force sensors and actuators for the study of cell signaling mechanics \cite{wang2013defining,kim2021double,brockman2018mapping,brockman2020live,ma2021dna, ma2019dna} and the control of DNA nanostructures \cite{gur2021double,lee2021characterizing}. 

\begin{figure}
    \centering
    \includegraphics[width=84mm]{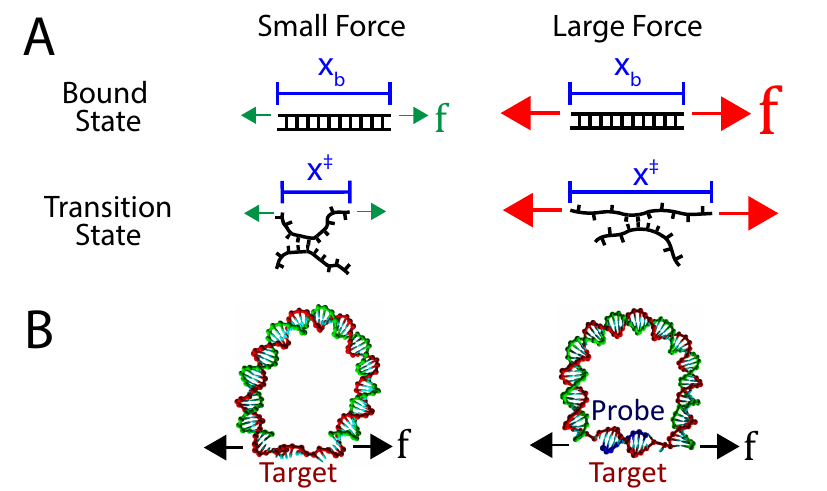}
    \caption{\textbf{(A)} A proposed model for how the extension of a duplex differs between small and large forces. The elasticity of the bound state is rigid, and therefore its extension $x_b$ is mostly unaffected by force. On the other hand, the transition state may be more flexible, in which case its extension $x^\ddagger$ will be force-dependent. In this case, worm-like chain models predict that at large forces, $x^\ddagger > x_b$, whereas at small forces $x^\ddagger < x_b$. \textbf{(B)} DNA bow assay: a bent bow-like duplex of variable length exerts tension on a \SI{15}{\nt} ssDNA target (bowstring), extending the strand.}
    \label{fig:dna_bow_model}
\end{figure}

In general, the force ($f$) dependence of two-state binding and unbinding kinetics can be modeled with a one-dimensional extension coordinate $x$ as \cite{dudko2008theory,guo2018structural}
\begin{equation}
    k_\alpha(f)=k_0 \exp\left(\int_0^f\Delta x^\ddagger(f^\prime)df^\prime/k_\mathrm{B}T\right),
    \label{eq:generalized_force_model}
\end{equation}
where $k_\alpha$ is the rate constant for binding ($\alpha=\mathrm{on}$) or unbinding ($\alpha=\mathrm{off}$), $\Delta x^\ddagger$ is the extension of the transition state ($x^\ddagger$) relative to the unbound ($x_\mathrm{ub}$) or bound state ($x_\mathrm{b}$), and $k_\mathrm{B} T$ is the thermal energy. If the transition state is more extended than the bound state by a constant ($\Delta x^\ddagger>0$), Equation~\ref{eq:generalized_force_model} yields the well-known Bell's formula \cite{bell1978models}: $k_{\mathrm{off}}\sim \exp(f\Delta x^\ddagger/k_\mathrm{B} T)$, which predicts that $k_{\mathrm{off}}$ monotonically increases with force. For DNA hybridization/dehybridization, the transition state is thought to be a nucleated duplex that contains both single-stranded DNA (ssDNA) and double-stranded DNA (dsDNA) \cite{vologodskii2018dna,porschke1971co,craig1971relaxation}. According to the worm-like chain model, ssDNA, whose persistence length ($A$) is $\sim1$ nm, behaves like a flexible chain in the low force regime ($f <k_\mathrm{B} T/A\sim \SI{5}{\pico\newton}$) \cite{camunas2016elastic}. It is thus conceivable that the transition state could be less extended than the pure dsDNA state in the low force regime (Figure~\ref{fig:dna_bow_model}A). Based on this idea, it was recently proposed that $k_{\mathrm{off}}(f)$ can decrease with force until $f\sim\SI{5}{\pico\newton}$ before increasing in the high force regime \cite{guo2018structural,guo2019understanding,wang2019force}. This counter-intuitive effect known as ``roll-over" was predicted in a recent single-molecule fluorescence-tweezers experiment \cite{whitley2017elasticity}, but the limited data leaves the conclusion in question. Furthermore, how the extension of the nucleated duplex in the transition state compares to that of dsDNA in the bound state and pure ssDNA in the unbound state is not known.  

\begin{figure}
    \centering
    \includegraphics[width=84mm,keepaspectratio]{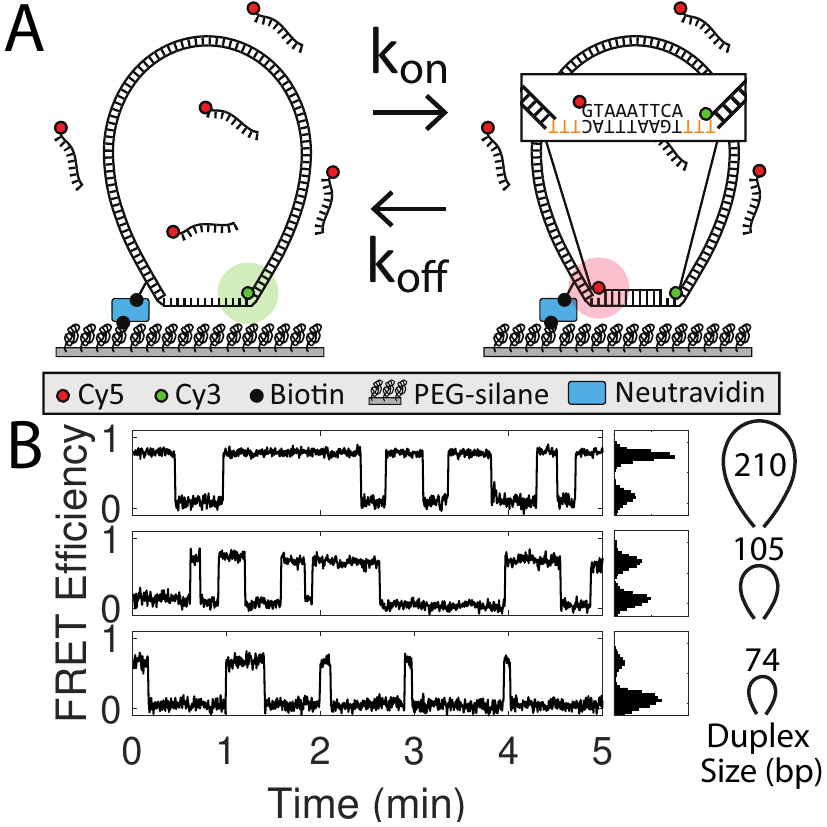}
    \caption{\textbf{(A)} Schematic of DNA bow assay FRET setup. Cy3-labeled DNA bows are immobilized on a PEGylated coverslip and excited by an evanescent wave of a 532-nm laser using TIRF microscopy. The inset highlights the ssDNA sequence (TGAAATTAC) targeted by the Cy5-labeled probe (GTAAATTCA). To avoid additional stacking interactions between the probe and the DNA bow, the \SI{9}{\nt} target segment was flanked by \SI{3}{\nt} ssDNA gaps (highlighted orange) in all construct designs. \textbf{(B)} Example FRET efficiency traces for three different dsDNA arc lengths (210, 105, 74 \SI{}{\bp}) exerting three separate forces (1.8, 3.8, 6.3 \SI{}{\pico\newton}, respectively). FRET histograms are shown right. Binding and unbinding rates are extracted from the mean dwell times of low and high FRET states respectively.}
\label{fig:dna_bow_fret}
\end{figure}

Here, we developed a DNA construct dubbed ``DNA bow" (Figure \ref{fig:dna_bow_model}B) to exert tension in the range between \SIrange{2}{6}{\pico\newton} on a short DNA oligo. The DNA bow is composed of a dsDNA segment (arc) of variable size ($\sim \SI{100}{\bp}$) and a short ssDNA target (bowstring); during experiment, a complementary ssDNA probe binds to and unbinds from this bow target. Combined with single-molecule FRET, DNA bows allow for high-throughput measurements of DNA hybridization and dehybridization kinetics in the low-force regime, using a conventional TIRF microscopy setup (Figure \ref{fig:dna_bow_fret}). Thus, this assay complements low-throughput, calibration-heavy tweezers \cite{whitley2017elasticity,shon2019submicrometer}. Using the DNA bow assay, we measured the hybridization and dehybridization rates of four DNA-DNA homoduplexes (lengths ranging from \SIrange{8}{9}{\bp}) as well as their corresponding RNA-DNA heteroduplexes. Overall, the measured dehybridization (unbinding) rate monotonically increased with force with no clear sign of roll-over, and the measured hybridization (binding) rate also increased with force. In agreement with these experimental results, our simulations reveal that hybridization and dehybridization of short oligos transition through a maximally extended state, and as a result both processes are accelerated in the low force regime. We attribute the extension of the transition state to steric repulsion, which prevents the ssDNA overhangs of the nucleated duplex from coiling.

\section{MATERIALS AND METHODS}

\subsection{Preparing DNA bows}
\begin{figure*} 
\centering
    \includegraphics[width=84mm,keepaspectratio]{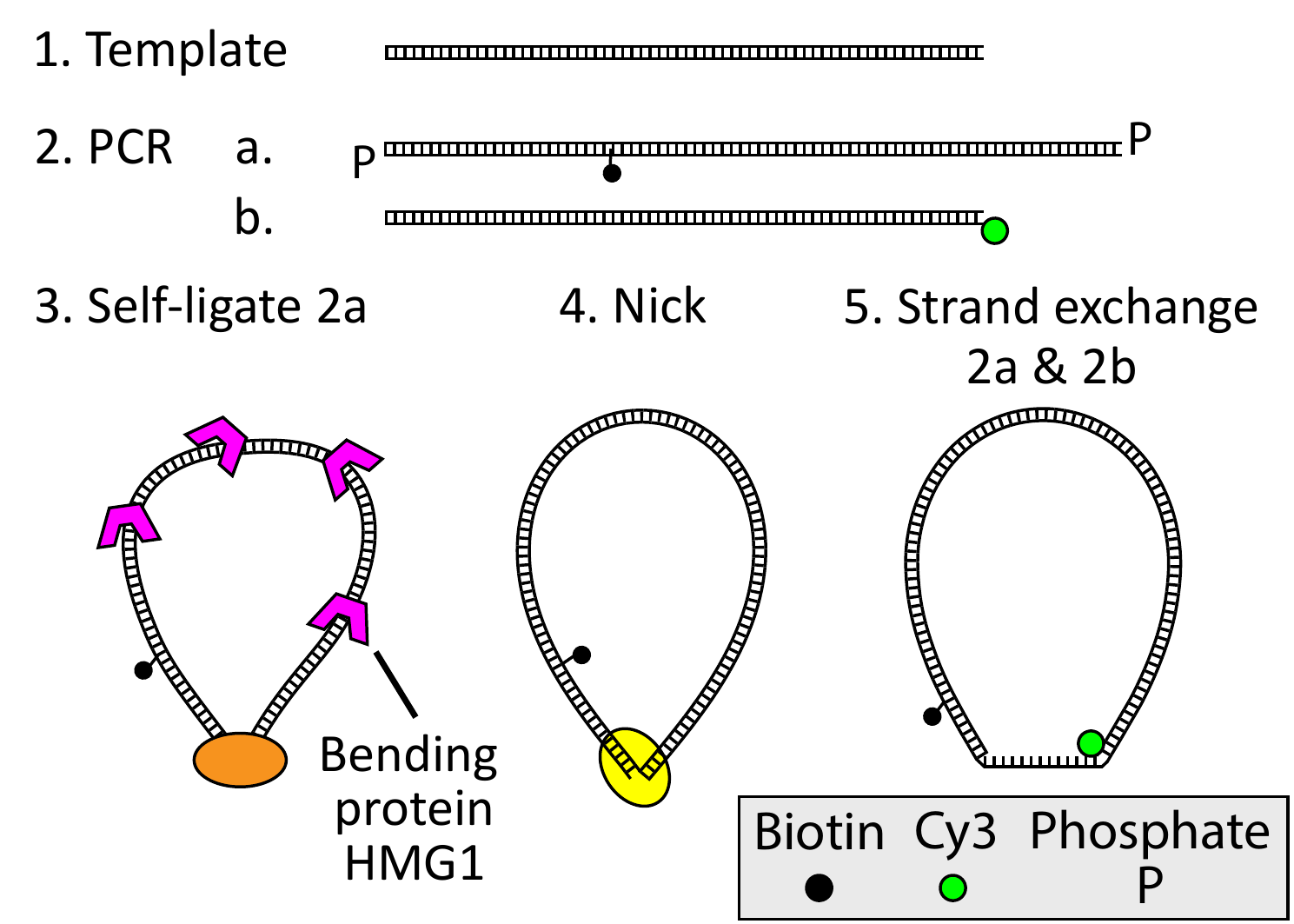}
    \caption{DNA bows were constructed in five steps. First, uniquely sized templates were generated with PCR from a common source. Using these templates, two sets of molecules (2a and 2b) were amplified with modified primers via PCR. DNA minicircles were then created from the phosphorylated 2a molecule set using protein-assisted DNA self-ligation. Afterward, DNA minicircles were purified and nicked on the unmodified strand. The final DNA bow constructs were finally constructed by exchanging the nicked strand of circularized 2a molecules with the Cy3-labeled 2b molecule.}
    \label{fig:dna_bow_construction}
\end{figure*}

DNA bow molecules were constructed and labeled with a FRET donor (Cy3) and a biotin linker in 5 steps (Figure \ref{fig:dna_bow_construction}): (1) Template generation, (2) Modifier incorporation, (3) Circularization, (4) Nick generation, and (5) Strand exchange. Most notably, DNA bending protein HMG1 was used to facilitate intramolecular ligation of short DNA molecules \cite{pil1993high}. 

In Step 1, polymerase chain reaction (PCR) was used to create a set of seven different DNA templates with lengths ranging from \SIrange{74}{252}{\bp} (Supplementary Figure \ref{sfig:configs_dna_bow_array}, Supplementary Table \ref{table:sequences}), using yeast genomic DNA as the source. The PCR primers were designed such that all seven templates shared adaptor sequences at their ends. In Step 2, using these templates, two additional PCR reactions were performed to create two sets of molecules, with each reaction using modified primers that anneal to the adaptor regions of the template. The first reaction produced a set of molecules with phosphorylated $5^\prime$ ends and an internal biotin-dT label for surface immobilization, as well as a \SI{15}{bp} extension, consisting of a \SI{9}{\bp} target segment flanked on both sides by (dT)$_{3}$ spacers. The second reaction produced donor-labeled (Cy3) molecules with a sequence identical to the original templates, which is \SI{15}{\bp} shorter than the first PCR product. All oligonucleotides were purchased from Eurofins MWC Operon and Integrated DNA Technology. All PCR products in the first and second steps were inspected by gel electrophoresis and extracted using a PCR clean-up kit. In Step 3, we circularized the phosphorylated molecules. To increase circularization efficiency, molecules were briefly incubated at \SI{15}{\nano \Molar} with \SI{0.75}{\micro \Molar} DNA bending protein HMG1 (Sigma Aldrich) in T4 ligase buffer for 10 minutes. Afterward, T4 ligase was added and the reaction volume was incubated overnight at \SI{15}{\celsius}. The reaction was stopped via heat inactivation, after which T5 exonuclease was added to remove linear inter-molecular or nicked intra-molecular ligation products. Finally, Proteinase K was added to remove any protein leftovers. The remaining circular molecules were purified and concentrated using ethanol precipitation. In Step 4, the unmodified strand of our circular molecules was nicked using Nb.BbvCI in 1x CutSmart buffer (NEB). After circularization and nicking, the resulting product was visualized and purified on a native polyacrylamide gel (6\%, 29:1 acrylamide to bis-acrylamide in 0.5x TBE buffer) , which appeared as a a single, isolated band as shown in Supplementary Figure \ref{sfig:mc_gels}. The bands were extracted using a simple ``crush-and-soak" method, and then concentrated using the same ethanol precipitation method as before. In Step 5, a strand-exchange reaction was performed, replacing the nicked strand on each circular molecule with the corresponding donor-labeled linear strand. Circular molecules were mixed with the donor-labeled linear molecules at a 4:1 ratio, briefly heated to \SI{95}{\celsius}, and gradually cooled down to \SI{4}{\celsius}.

\subsection{DNA bow assay}
Microscope slides with pre-drilled holes and coverslips were cleaned by sonicating in deionized water, drying in a vaccuum chamber, and 5-minute etching in a plasma chamber. The cleaned slides and coverslips were then passivated with PEG (polyethylene glycol) to minimize nonspecific binding. After PEGylation, the flow cell was assembled by joining the slide and the coverslip with double-sided tape and epoxy glue. The flow cell interior was incubated with NeutrAvidin followed by \SI{50}{\micro\liter} of \SI{40}{\pico \Molar} DNA bow solution. Each measurement began after perfusing \SI{20}{\nano\Molar} of ssDNA probe solution into the flow chamber. The temperature of the flow chamber was maintained at \SI{22}{\celsius} using an objective lens temperature controller. For each molecule, a high Cy3 signal (low FRET) indicates a DNA bow in the unbound state, while a high Cy5 signal (high FRET) indicates a DNA bow bound with the probe (Figure \ref{fig:dna_bow_fret}). Bound and unbound lifetimes of approximately $\sim100$ immobilized molecules were collected in each trial; 2-4 trials were performed for each bow size. All data was collected on an objective-based TIR microscope with an EMCCD camera (DU-897ECS0-BV, Andor). Frame times varied from \SIrange{50}{1000}{\milli \second}, depending on the duplex sequence. The imaging buffer contained \SI{100}{\milli\Molar} NaCl, \SI{100}{\milli\Molar} Tris (\SI{8}{\pH}), a triplet state quencher (1 mM Trolox), and the protocatechuic acid (PCA)/protocatechuate-3,4-dioxygenase (PCD) system \cite{aitken2008oxygen}. Using this system, photobleaching was negligible at all donor excitation power settings and camera acquisition times used in our experiments.

\subsection{Data analysis}
For each trial, time trajectories of FRET values were extracted from surface-immobilized molecules with in-house Matlab codes. Briefly, we calculated the FRET signal for each molecule from the background-subtracted intensities of the donor signal ($I_D$) and the acceptor signal ($I_A$) with $I_A/(I_A+I_D)$. Next, we filtered FRET trajectories with a moving average, and used FRET signal thresholding to mark discrete transitions between the two FRET states. The dwell times in the bound (``on") state (high-FRET state) and the unbound (``off") state (low-FRET state) were collected from each FRET trajectory. The binding rate ($k_\mathrm{on}$) and the unbinding rate ($k_\mathrm{off}$) were calculated from the mean dwell times ($\tau$) using $k_\mathrm{on}=([\mathrm{c}]\tau_\mathrm{off})^{-1}$ and $k_\mathrm{off}=\tau_\mathrm{on}^{-1}$, where $[\mathrm{c}]$ is the concentration of Cy5 labeled probes. Typically, $\sim150$ trajectories were used for each rate measurement.

\subsection{Estimating the tensile force exerted by a DNA bow}

To estimate the tension exerted on the ssDNA bowstring, we treated the dsDNA arc as a worm-like chain. The force $f$ exerted by a worm-like chain along its end-to-end direction at distance $x_0$ can be calculated from the end-to-end distance ($x$) distribution $P(x)$ of the chain according to
\begin{equation}\label{eq:tensile_force}
    f(x_0) = -k_BT\frac{\partial \log P(x)}{\partial x}\Big|_{x_0}.
\end{equation}
For $P(x)$, we used an interpolated formula (Supplementary Equation \ref{eq:radial_distribution}), which is accurate for a wide range of bending stiffness values \cite{becker2010radial}. 

With our bow design, $x_0$ also corresponds to the equilibrium extension of the ssDNA bowstring, and therefore its value will depend on both the bow size as well as whether the probe is bound to the complementary target segment. To find a realistic value of $x_0$, we performed oxDNA2 simulations \cite{snodin2015introducing,vsulc2012sequence,gravina2021coarse} for all possible combinations of bow size, target sequence, and probe state (bound or unbound). DNA bows bound to an RNA probe were not simulated; while oligomeric RNA-DNA duplexes have a slightly smaller helical rise \cite{shaw2008recognition}, the overall effect that this difference would have on the force is negligible. Each MD simulation was run for $t=\SI{1.14}{\micro \second}$, using a time step of $\SI{15.2}{\femto\second}$. For each trajectory $n= 7.5\times10^4$ configurations were saved in \SI{15.2}{\pico\second} evenly spaced intervals. Using these saved configurations, we calculated the extension $x$, defined as the distance between the bases located at the terminal ends of the dsDNA bow and linked to the ssDNA target strand. The exact location of each terminal base was specified by its center of mass. Afterward, we calculated the mean extension ($\overline{x}$) and standard deviation $\sigma(x)$ for each molecule's $x$ distribution, to estimate $x_0$ and its associated uncertainty respectively (Supplementary Table \ref{table:e2e_distances}, Supplementary Figure \ref{sfig:dna_bow_distributions}). The tensile force $f(\overline{x})$ was then calculated using Equation \ref{eq:tensile_force}, and the uncertainty in the tensile force was estimated with $\sigma(x)$ by propagation of error, using $\partial f(x)/\partial x\Big|_{\overline{x}}\cdot \sigma(x)$. Additional details regarding WLC parameters and oxDNA2 simulations are provided in Supplementary Materials.

\subsection{Estimating hybridization and melting rates with FFS simulations}
 
To determine hybridization and melting rates, we used a technique known as ``forward flux sampling" (FFS) \cite{allen2005sampling,allen2009forward}. This method ratchets the rare transition from an unbound state to a bound state, or vice versa, using a series of checkpoint interfaces which are each characterized by a unique order parameter value (e.g. minimum distance, number of bonds). By measuring the average flux $\Phi_{A,0}$ of a molecule in state A crossing the first interface $\lambda_0$, and then measuring the probability $P(\lambda_{Q}|\lambda_{Q-1})$ of transitioning from $\lambda_{Q-1}$ to $\lambda_Q$ at each subsequent interface, it is possible to estimate the overall transition rate to state B ($k_{AB}$ ), according to:
\begin{equation}\label{eq:ffs}
    k_{AB} = \Phi_{A,0}\prod_{Q=1}^{n}P(\lambda_{Q}|\lambda_{Q-1}).
\end{equation}
Using this technique with oxDNA2, the rate of the probe P1-DNA binding to or unbinding from its corresponding target sequence T1 was calculated (Supplementary Table \ref{table:sequences}). Additional simulation details and parameter values are provided in Supplementary Materials.


\subsection{Observing the force-extension behavior of near-transition oligoduplexes}
 To measure the force-extension behavior of a partially melted oligoduplex near its binding or unbinding transition, we performed a series of MD simulations using the ``mutual trap" external force tool provided with oxDNA2. Similar to the previous section, we simulated the target strand in four states: the ``probe-bound" state (\SI{9}{\bp}), the ssDNA ``probe-unbound" state (\SI{0}{\bp}), a transition state with \SI{1}{\bp} remaining at the $3^\prime$ end of the duplex, and a transition state with \SI{1}{\bp} remaining at the center. In the transition state simulations, the remaining terminal or middle base pair interaction was strengthened 10-fold, while all other base pairing interactions were set to zero. For all simulations, the ends of the target strand were connected by a harmonic spring with stiffness $k = \SI{57.1}{\pico\newton\per\nano\meter}$ (1 simulation unit) and relaxed extension $x_0$, such that the the tension $f$ and extension $x$ of the strand could easily be related using $f=-k\cdot(\overline{x}-x_0)$. Similar to our DNA bow simulations, the extension $x$ was defined as the distance between the center of mass of each terminal base on the target strand. For each state, we performed MD simulations for a small range of $x_0$ values, such that the corresponding forces approximately spanned the force range of our DNA bows. For comparison, we plot the force-extension behavior of the target strand extended by a harmonic spring or a DNA bow in Supplementary Figure \ref{sfig:spring_vs_bow}. Each simulation was performed for $t=\SI{1.52}{\micro \second}$ using a time step of $\SI{15.2}{\femto\second}$. $n=10^5$ pairs of force and extension values were then calculated from configurations collected in \SI{15.2}{\pico\second} intervals evenly spaced across the MD trajectory.

\section{Results and Discussion}

Using the DNA bow assay, we measured the binding and unbinding rates of a short DNA or RNA (8- or 9- \SI{}{\nt}) oligo to a weakly pulled complementary target strand (\SI{15}{nt}). The measured binding ($k_\mathrm{on}$) and unbinding ($k_\mathrm{off}$) rate constants thus reflect hybridization and dehybridization transitions of a short DNA homoduplex or DNA/RNA heteroduplex. Our DNA bow assay exploits the bending rigidity of dsDNA to generate small forces and is conceptually similar to the force clamp implemented with DNA origami \cite{nickels2016molecular} and a loop-based force transducer  \cite{mustafa2018force}. An identical DNA construct has also been used in other studies \cite{shroff2005biocompatible,kim2015dynamic}. Our DNA bow assay offers unique advantages over other single-molecule force assays such as optical and magnetic tweezers in that (1) force measurements can be performed on many molecules in parallel, and (2) no calibration of force vs. extension is required for each molecule since the force is generated by chemically identical DNA molecules, not through beads of variable properties. We created 21 DNA bows in total, including 7 different dsDNA lengths (74, 84, 105, 126, 158, 210 and \SI{252}{bp}) for the elastic arc segment and 3 unique sequences for the complementary segment of the ssDNA target. The DNA bow was further designed such that only the desired gapped DNA circle can generate the FRET signal from the surface upon probe binding (Supplementary Figure \ref{sfig:dna_bow_table}).  While sharp bending is known to disrupt the helical structure of circular DNA by generating ``kinks", these deformations do not appear in circles larger than \SI{84}{\bp} \cite{du2008kinking}. Therefore, we predict that kinking is negligible even for our smallest DNA bow size, which includes a flexible \SI{15}{\bp} ssDNA segment in addition to its \SI{74}{\bp} dsDNA arc. Given this, the force generated by each DNA bow was calculated by treating the DNA arc as a simple worm-like chain. Using this assumption, the range of forces exerted on the target strand is calculated to be \SIrange{1.70}{6.34}{\pico\newton} in the unbound state and \SIrange{1.6}{6.25}{\pico\newton} in the bound state.

\begin{figure}
    \centering
    \includegraphics[width=\textwidth]{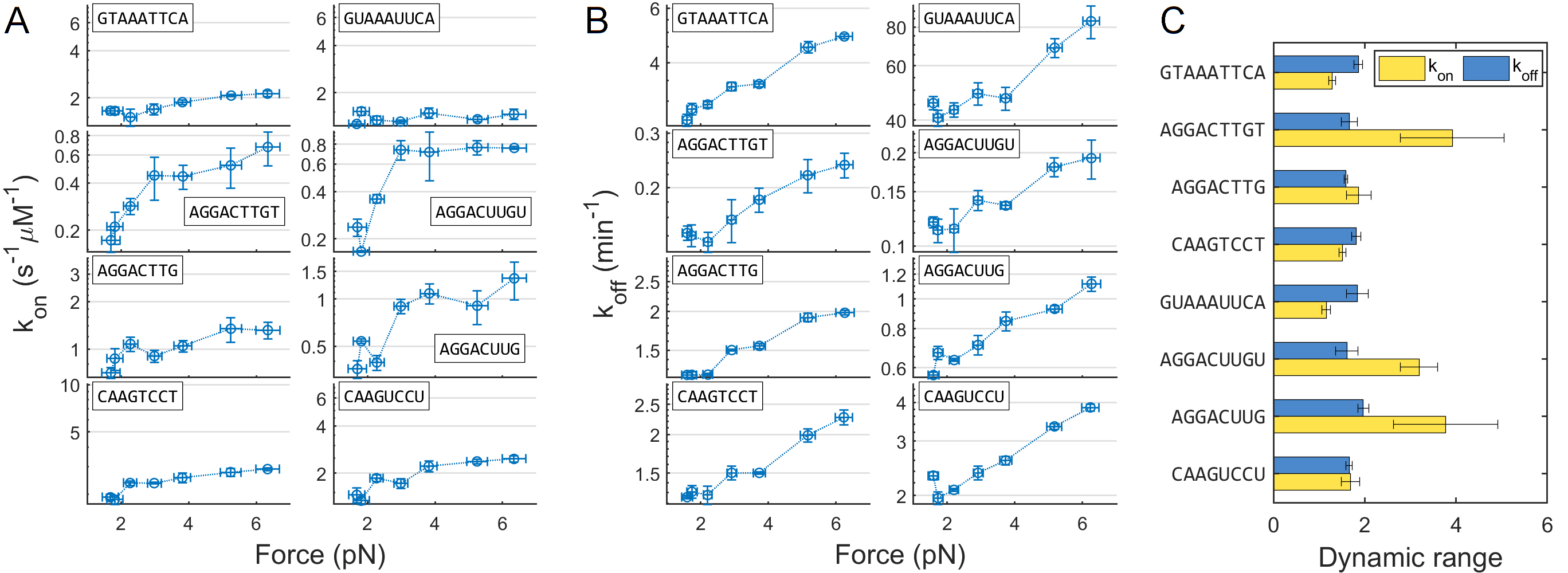}
    \caption{\textbf{(A)} Binding rate vs. force. The plots on the left (right) column are for DNA (RNA) probes. The y-axis is on a logarithmic scale over the same 5.8-fold change for all probe sequences. \textbf{(B)} Unbinding rate vs. force. The plots on the left (right) column are for DNA (RNA) probes. The y-axis is on a logarithmic scale over the same 2.4-fold change for all probe sequences. Vertical error bars for binding and unbinding rates represent the standard error of the mean; horizontal error bars were calculated using $\frac{\partial f(x)}{\partial x}\Big|_{\overline{x}} \cdot \sigma(x)$, where $\overline{x}$ and $\sigma(x)$ are the mean and standard deviation of the bow's end-to-end distance distribution. \textbf{(C)} The dynamic range of all measured rates. The dynamic range was obtained by dividing the rate at the highest force by that at the lowest force; the associated error was calculated by propagating the uncertainty in the underlying rates.}
    \label{fig:alldata}
\end{figure}

\subsection{Binding and unbinding rates vs. force}

In Figure \ref{fig:alldata}, we present the force dependence of $k_\mathrm{on}$ (A) and $k_\mathrm{off}$ (B) for 4 DNA-DNA duplexes (left column) and 4 RNA-DNA duplexes (right column). The scale of y-axis is set as logarithmic to aid comparison to Equation~\ref{eq:generalized_force_model}.  Each RNA sequence is identical to a corresponding DNA sequence, except for T to U substitution. As shown in Figure \ref{fig:alldata}A, $k_{\mathrm{on}}$ tends to increase with force over the measured force range. The relative increase in $k_{\mathrm{on}}$ is sequence-dependent: the increase is relatively large for AGGACTTGT but small for GTAAATTCA. The relative increase or dynamic range is quantified by taking the ratio of the rate at the highest force to that at the lowest force (Figure \ref{fig:alldata}C). This sequence-dependence was also observed in RNA-DNA duplexes, with each heteroduplex approximately matching the behavior of its corresponding homoduplex. However, these differences in relative increase mostly disappear above $\SI{3}{\pico\newton}$, and $k_{\mathrm{on}}$ appears to reach a plateau above \SI{6}{\pico\newton}. We note that the comparison of the second-order rate constant $k_{\mathrm{on}}$ across different sequences is not accurate because of the inaccuracy in the estimated concentration of each probe.  

The most significant result from Figure \ref{fig:alldata}A is that $k_{\mathrm{on}}$, the binding rate of the probe to its complementary target, becomes faster, not slower, as the tension in the target strand increases. This result stands in contrast to previous rates observed at higher forces, such as those observed for DNA hairpin folding \cite{woodside2006nanomechanical,liphardt2001reversible, alemany2017force}.
By differentiating the logarithm of Equation~\ref{eq:generalized_force_model} with respect to force, we can relate the slope of curves in Figure \ref{fig:alldata}A to $\Delta x^\ddagger$, which is the extension of the transition state ($x^\ddagger$) relative to the unbound state ($x_\mathrm{ub}$):
\begin{equation}\label{eq:derivative_kon}
    \frac{d\log{k_{\mathrm{on}}(f)}}{df}=\frac{\Delta x^\ddagger(f)}{k_B T}.
\end{equation}
The overall non-negative slope in Figure~\ref{fig:alldata}A indicates that the transition state for hybridization is more extended than the unbound state ($x^\ddagger>x_\mathrm{ub}$) in the range of \SIrange{2}{6}{\pico\newton}.

The force dependence of $k_{\mathrm{off}}$ is shown in Figure \ref{fig:alldata}B. Compared to $k_{\mathrm{on}}$, the dynamic range for $k_{\mathrm{off}}$ is somewhat uniform at 2-fold across all DNA-DNA and DNA-RNA duplexes (Figure~\ref{fig:alldata}C). The apparent slope is mostly positive except between a few points below \SI{2}{\pico\newton}, which implies that the roll-over effect or catch-to-slip transition is negligible. Similar to Equation~\ref{eq:derivative_kon}, the slope of curves in Figure \ref{fig:alldata}B is proportional to $\Delta x^\ddagger$ for dehybridization, which is the extension of the transition state ($x^\ddagger$) relative to the bound state ($x_\mathrm{b}$). From this, we conclude that the transition state for dehybridization is more extended than the bound state ($x^\ddagger>x_\mathrm{b}$) in the range of \SIrange{2}{6}{\pico\newton}.   

Since the first-order rate constant $k_{\mathrm{off}}$ is concentration-independent, it can be compared across different sequences. When compared at the same force, $k_{\mathrm{off}}$ was in the order of GTAAATTCA $>$ AGGACTTG $=$ CAAGTCCT $>$ AGGACTTGT from fastest to slowest. When a single nucleotide was removed from the $3^\prime$ end of AGGACTTGT, $k_{\mathrm{off}}$ increased as expected from the weaker base pairing interaction. Between AGGACTTG and its reverse complement CAAGTCCT, $k_{\mathrm{off}}$ remains the same, which implies that for a DNA-DNA homoduplex, $k_{\mathrm{off}}$ is similar regardless of which strand is subject to tension. $k_{\mathrm{off}}$ for RNA-DNA duplexes (Figure \ref{fig:alldata}A, right) similarly showed a strong sequence-dependence, in the order of GUAAAUUCA$>$CAAGUCCU$>$AGGACUUG$>$AGGACUUGU. In two cases (AGGACUUGU and AGGACUUG), RNA-DNA heteroduplex was longer-lived than its homoduplex counterpart, but in the other two (GUAAAUUCA and CAAGUCCU), DNA-DNA homoduplex was longer-lived. 

\subsection{Thermodynamic stability}
From the individual rate constants, we can calculate the standard free energy difference ($\Delta G^\circ$) between the bound and unbound states according to
\begin{equation}
    \Delta G^\circ = k_\mathrm{B}T \log\frac{k_\mathrm{on}[c_0]}{k_\mathrm{off}}
\end{equation}
where $[c_0]$ is \SI{1}{\molar}. In this definition, $\Delta G^\circ$ is more positive for a more stable duplex. In Supplementary Figure \ref{sfig:NN_barplot}, we compare $\Delta G^\circ$ calculated using $k_\mathrm{on}^0$ and $k_\mathrm{off}^0$ with $\Delta G_\mathrm{NN}^\circ$ estimated using a nearest-neighbor (NN) thermodynamic model \cite{santalucia1998unified}. Most sequences are significantly more stable than the model predicts, showing at least a $2\ k_B T$ difference. This increased stability can be attributed to two major factors. First, the terminal bases of the duplex will stack with the adjacent unpaired bases in the gaps, which has been shown to provide $\sim 1\ k_B T$ per end interaction in \SI{8}{\bp} DNA duplexes \cite{bommarito2000thermodynamic}. Dangling nucleotides beyond these adjacent bases have also been shown to stabilize the duplex \cite{doktycz1990thermodynamic,senior1988influence}, albeit to a lesser degree \cite{santalucia2004thermodynamics}. Second, the DNA and RNA probes used in this experiment were labeled with a Cy5 dye on the $5^\prime$ end, which will also stabilize short DNA duplexes by $2\ k_B T$ \cite{moreira2015cy3}. The stabilizing effects of dangling-base interactions and $5^\prime$ dye labeling are additive \cite{moreira2015cy3}. When accounting for these two factors, we find that the nearest-neighbor model prediction matches $\Delta G^\circ$ more closely.

In Supplementary Figure~\ref{sfig:G_vs_f}, the free energy difference $\Delta G^\circ$ is plotted against force. Because both rates change in the same direction in response to force, the force dependence of $\Delta G^\circ$ is somewhat dampened. Except for AGGACTTGT and its RNA counterpart, $\Delta G^\circ$ changes little, albeit with some scatter. In comparison, the force-dependence of $\Delta G^\circ$ of AGGACTTGT and AGGACUUGU shows a monotonic increase up to \SI{3}{\pico\newton} and afterward plateaus, varying by less than $0.5\ k_B T$.

\subsection{oxDNA2 simulations of binding and unbinding trajectories}
To gain molecular insights into binding and unbinding transitions, we performed coarse-grained simulations of both reactions using oxDNA2. For both simulations, the probe P1-DNA was simulated together with its corresponding target sequence T1 (Supplementary Table \ref{table:sequences}). The end-to-end extension of the target strand was held fixed using the harmonic trap tool provided with oxDNA2, while the probe was allowed to diffuse freely. The target strand was held at \SI{5.5}{\nano\meter} and \SI{5.1}{\nano\meter} for unbinding and binding reactions, respectively. These values were determined from the average target strand extensions of the largest DNA bow in the bound and unbound states. Because binding and unbinding events are relatively rare, we implemented forward flux sampling, which separates rare events into computationally feasible intervals \cite{allen2009forward,allen2005sampling}. Each interval was demarcated by two interfaces, where each interface was defined using a relevant order parameter (Supplementary Tables \ref{table:binding_order_parameters} and \ref{table:unbinding_order_parameters}). Using Equation \ref{eq:ffs}, we calculated an average $k_\mathrm{on}$ value of \SI{2.3}{\per\second\per\micro\molar} (where the probe concentration was estimated using the $[\SI{10.2}{\nano\meter}]^3$ simulation box volume) and an average $k_\mathrm{off}$ value of \SI{9.9}{\per\min}.  A direct comparison between the calculated rates and the measured rates is not accurate considering that coarse-graining is known to speed up dynamical timescales by smoothing energy landscapes and neglecting hydrodynamic effects \cite{sengar2021primer, murtola2009multiscale, guenza2015thermodynamic}. Nonetheless, the reaction paths should shed light on the nature of the transition states. 


\begin{figure}
    \centering
    \includegraphics{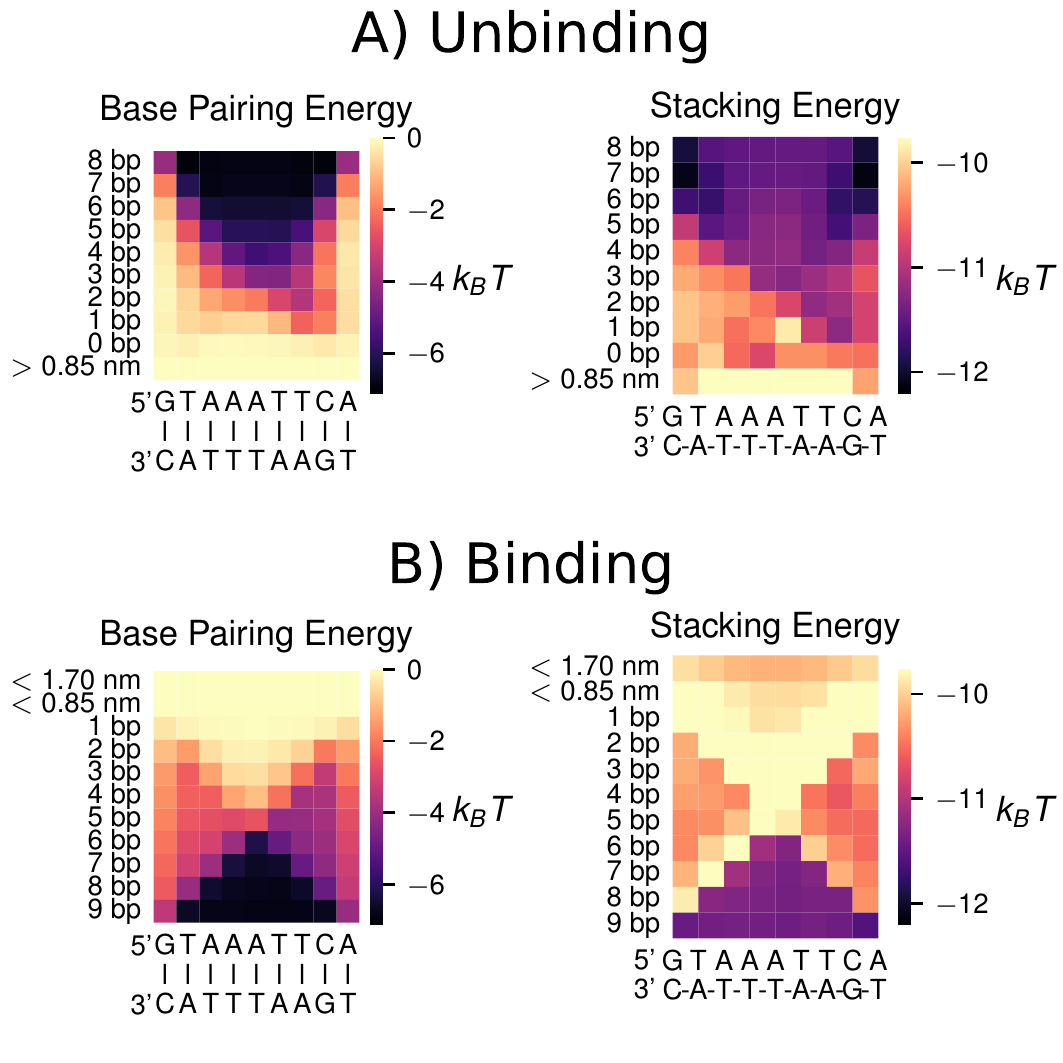}
    \caption{Average pairing and stacking potentials for the complementary portion of the target strand at each interface of \textbf{(A)} binding and \textbf{(B)} unbinding FFS simulations. Colorbar ticks are specified in $k_BT$ units, where T = \SI{22}{\celsius}. }
    \label{fig:binding_unbinding_heatmaps}
\end{figure}

The initial unbinding rate of the first-melted base, as well as the melting probability of each subsequent base, are individually tabulated in Table \ref{table:ffs_results}. An additional step is also included which calculates the probability that the strands exceed a minimum distance $d=\SI{0.85}{\nano\meter}$ after all bases have melted (Table \ref{table:unbinding_order_parameters}). Throughout unbinding, the probability of melting each base is relatively small, ranging from 0.03 to 0.06. Notably, the chance that oligos separate from one another remains low even after all bases have melted: according to the final strand separation step, a newly-melted duplex is expected to re-form a base pair with $p\sim0.9$ probability. This fast reassociation between short oligos is similar to that recently observed between DNA and the lac repressor \cite{marklund2022sequence}. This result also suggests that the unbinding transition state happens after all base pairs have already melted. Therefore, the apparent transition state for a FRET-based dissociation event should occur after the oligos are physically well separated by some distance.

We also measured the average base-pairing and stacking potentials of the complementary target segment for all unbinding steps (Figure \ref{fig:binding_unbinding_heatmaps}A). As the unbinding reaction progressed, both pairing and stacking potentials weakened in a symmetric fashion; interactions on the edges of the duplex were more likely to be broken than interactions nearest to the center. Stacking interactions were largely unaffected during the first 3 melting steps ($> \SI{6}{\bp}$), and remain relatively high up to the final base melting step ($\SI{0}{\bp}$). During the last few steps (\SIrange{3}{0}{\bp}), the remaining interactions of GTAAATTCA are skewed towards one side in an apparent symmetry breaking, centered on the relatively strong CT dinucleotide pair. The discrete change in the stacking potential after strand separation ($d>\SI{0.85}{\nano\meter}$) suggests that the oligos maintain some residual helical stacking immediately after duplex melting, which may enable the fast reassociation previously discussed.

Similar to unbinding, the individual steps of binding are also enumerated in Table \ref{table:ffs_results}. They include two strand approach steps (where interfaces are defined by inter-strand separation going below a minimum distance threshold) as well as nine base pairing steps (Table \ref{table:binding_order_parameters}). The slowest step in the binding process was the formation of the first base pair (starting at minimum distance $d = \SI{0.85}{\nano\meter}$ between matching bases). The success probability of this step was less than $5\times10^{-3}$, an order of magnitude lower than in any other step. After this step, however, the probability of additional base pairing increases rapidly, in stark contrast to the low probabilities seen throughout the unbinding reaction. The likelihood of full duplex formation approaches one after only four bases have formed, consistent with the zipping model for DNA binding \cite{applequist1963theory,gibbs1959statistical}. 
 
As before, we measured the pairing and stacking potentials of both sequences as binding progressed (Figure \ref{fig:binding_unbinding_heatmaps}B). In stark contrast to unbinding, all pairing potentials first strengthen at either end of the target strand and afterwards ``zip" in a linear fashion. While both pathways were common, we observed a slight $5^\prime$ to $3^\prime$ preference for the target strand zipping direction. Stacking potentials increased in strength in a similar fashion; however, these potentials temporarily weaken just before pairing occurs. This result suggests that local unstacking may promote duplex nucleation by granting more orientational freedom to bases. 

Comparing these reactions to unbinding, we find that hybridization is not a simple reversal of melting (Supplementary Figure \ref{sfig:DNA_reaction_flowchart}). Melting adopts a ``fray and peel" pathway, where unbinding begins at the duplex termini and slowly proceeds base by base toward the duplex center. By contrast, hybridization follows a ``dock and zip" pathway, where the strands anneal with high probability after the rare formation of a toehold at a strand terminus. Together, these differences demonstrate that binding and unbinding take different reaction pathways and do not share the same transition barrier.

\subsection{The physical nature of the transition state(s)}
\begin{figure}
    \centering
    \includegraphics[width=\textwidth]{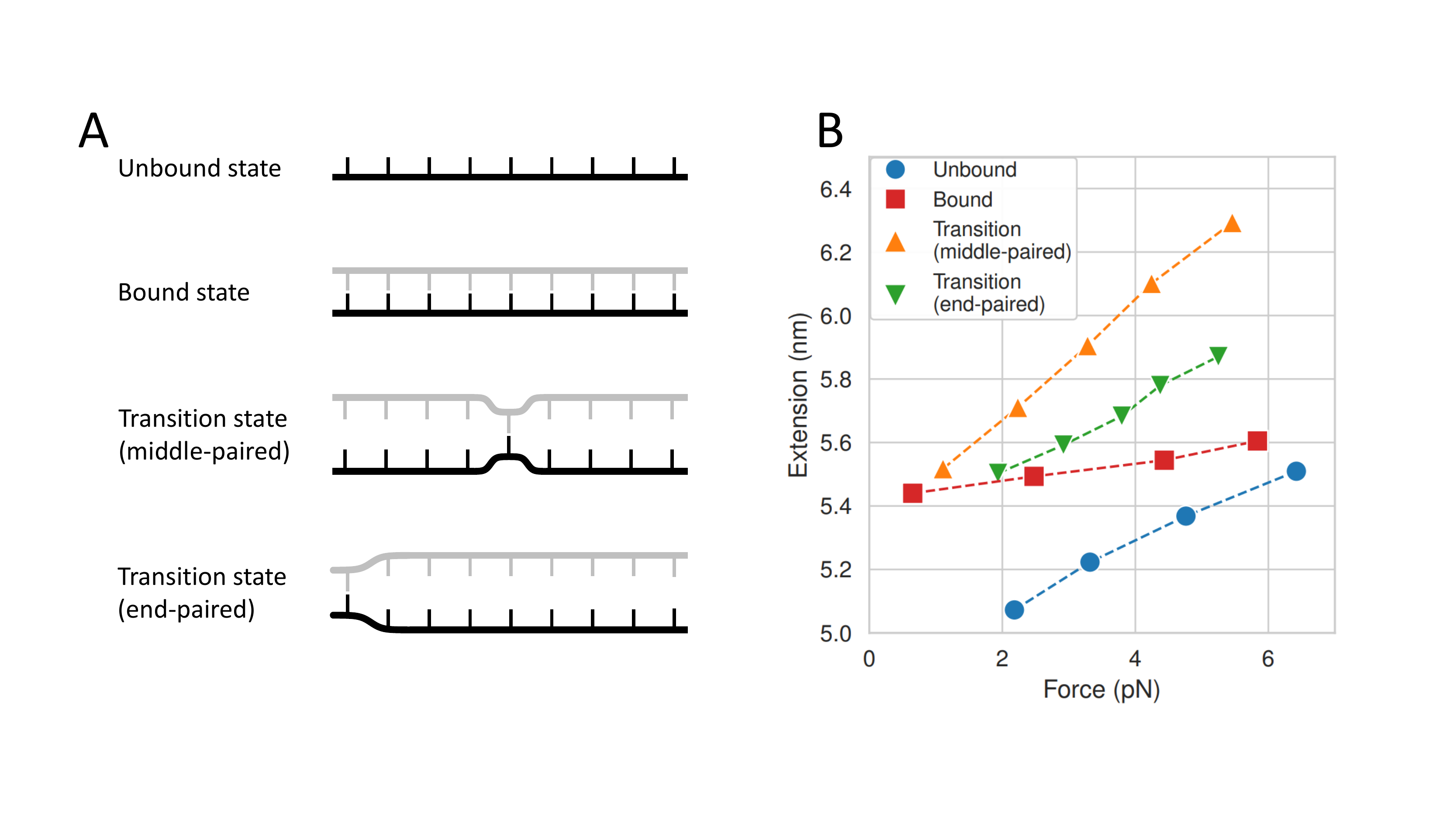}
    \caption{\textbf{(A)} Schematic of target strand (black) in four unique states: the unbound state, the bound state, the transition state with one base pair in the middle, and the transition state with one terminal base pair. \textbf{(B)} Force-extension curves of target strand in each state. For each filled marker, a unique MD simulation of the target segment was performed ($t= \SI{75.75}{\nano\second}$, $dt = \SI{15.2}{\femto\second}$). Afterward, the extension value was estimated by calculating the average distance $\overline{x}$ between the center-of-masses of the terminal bases on the target strand, and force values were calculated using $f = -k\cdot(\overline{x}-x_0)$. Dotted lines represent a linear regression of force-extension values observed for state. For comparison, the average extension values and corresponding force values of all bow sizes containing the same target sequence are also shown, in both its bound and unbound states.}
    \label{fig:force_extensions}
\end{figure}

Our DNA-bow experiments show that in the force range of \SIrange{2}{6}{\pico\newton}, both the binding and unbinding rates increase with force. The fact that weak force increases the accessibility of the transition state implies that the transition state is more extended than the two observable states, bound and unbound. At the same time, our kinetics simulations show that hybridization and dehybridization do not share the same transition state. Hybridization more likely occurs through the formation of a few terminal base pairs (end-paired transition state) while dehybridization occurs after the breakage of the last remaining base pair in the center (middle-paired transition state). To rationalize our experimental results with extension $x$ as the sole reaction coordinate, we obtained the force-extension curves of bound, unbound, and two different transition states (Figure~\ref{fig:force_extensions}A) from oxDNA2 simulations. Since the transition state is too transient to be analyzed in a normal dynamics simulation, we stalled the system near this state by turning off all base pairing interaction except in one central or terminal base pair, whose pairing interaction was strengthened 10-fold. As shown in Figure~\ref{fig:force_extensions}B, we find that the transition state for hybridization (end-paired) is more extended than the unbound state (ssDNA), and the transition state for dehybridization (middle-paired) is more extended than the bound state (dsDNA) over the entire force range of our experimental assay. Hence, our simulation results are consistent with the measured force-dependence of both $k_\mathrm{on}$ and $k_\mathrm{off}$.  

At first sight, it is not obvious why the transition state, which is a mixed state of ssDNA and dsDNA, is more extended than the bound and unbound state, which are pure dsDNA and ssDNA, respectively. We speculate that ssDNA strands confined to close proximity prevent each other from adopting randomly coiled conformations. To the same effect, randomly coiled conformations of ssDNA strands are not compatible to form a nucleated duplex. Therefore, ssDNA regions in the transition state happen to be more extended than in isolation.   


We find that the higher extension of DNA in its transition state is likely created by exclusion interactions between the target and probe strands. When comparing the unbound state to the binding transition state, bases nearby the end pair location were on average much further from bases on the opposite side of the strand, suggesting that the presence of the probe in the binding transition state blocks folded conformations. This may explain why weak tension increases the accessibility of the transition state, which extends ssDNA without overstretching, bringing the target strand closer to its ``unfolded" transition state. For the unbinding transition, melted bases at one end of the complementary segment were much further from bases at the opposite end. This increased distance between the ssDNA overhangs occurs in spite of their increased flexibility, presumably due to exclusion interactions that occur between the target and the probe. 


\subsection{Comments on roll-over or catch-to-slip transition}

The roll-over effect was postulated based on the idea that the transition state is a hybrid of ssDNA and dsDNA and that each obeys the force-extension formula for an ideal WLC \cite{guo2018structural,wang2019force} or a unique WLC with its own characteristics \cite{whitley2017elasticity}. The interpolation formula used in these models, however, is not valid for short chains \cite{marko1995stretching,bouchiat1999estimating}. For example, the formula predicts that the average end-to-end distance of a 10-nt ssDNA or dsDNA is zero, which is obviously incorrect. A more accurate formula derived for short chains \cite{keller2003relating,hori2007stretching} places the crossover force at $\sim\SI{1.8}{\pico\newton}$ (Supplementary Note), which borders the force limit of our DNA bow assay. However, even this formula cannot accurately describe the force-extension behavior of a nucleated duplex, whose unpaired regions are unavoidably influenced by exclusion interactions. Instead, we used oxDNA2 simulations to directly obtain the force-extension curves of the short ssDNA, dsDNA, and transition state. As shown in Figure~\ref{fig:force_extensions}B, the transition state is more extended than either ssDNA or dsDNA across the entire force range of our experimental assay. However, our study does not completely eliminate the possibility of a roll-over. First, our DNA bow assay cannot probe forces lower than \SI{1.5}{\pico\newton}. In this range, we find that the extension of the transition state can become shorter than that of the bound state (Figure \ref{fig:force_extensions}). Second, the roll-over effect is predicted to be more pronounced for longer oligos \cite{wang2019force}. A more thorough test of this model thus requires measuring the dehybridization rate of oligos longer than \SI{10}{\nt}, which is extremely slow ($\sim\mathrm{hr}^{-1}$). Therefore, the roll-over effect, if any, would only exist on a time scale too slow to bear any physiological or practical significance beyond the theoretical realm. 

\section{CONCLUSION}
DNA often experiences tension through passive or active mechanisms. In the presence of \SI{5}{\pico\newton} of force, DNA polymer models predict that dsDNA and ssDNA have a similar extension, which can lead to a nontrivial force dependence of hybridization and dehybridization rates. Previous force spectroscopy techniques, however, are not suitable for investigating this force dependence due to limited throughput. In this study, we developed a DNA bow assay, which can exert \SIrange{2}{6}{\pico\newton} of tension on a ssDNA target and report on its hybridization and dehybridization via smFRET. In this force range, we found that both the hybridization and dehybridization rates increase with force, which indicates that the transition state has a longer extension than its ssDNA and dsDNA counterparts. Coarse-grained simulations reveal that hybridization and dehybridization proceed through different transition states with a single base pair formed near the end or the middle. Consistent with the experimental results, simulations also show that these two transition states are indeed more extended than their respective initial states due to exclusion interactions that preclude ssDNA overhangs from adopting random coil configurations. Our study underscores the importance of investigating DNA-based reaction kinetics in the low force regime, which are not predictable by canonical polymer models of DNA.

\section{DATA AVAILABILITY}
All data presented in this manuscript can be made available upon request from the corresponding author.

\section{ACKNOWLEDGEMENTS}
The authors thank the members of the Kim laboratory for useful discussions. Computational resources were provided by the Partnership for an Advanced Computing Envrironment (PACE) at the Georgia Institute of Technology.

\section{FUNDING}
National Institutes of Health [R01GM112882]. Funding for open access charge: National Institutes of Health.
\subsubsection{Conflict of interest statement.} None declared.
\newpage

\bibliography{references}

\clearpage
\newcommand{\beginsupplement}{%
        \setcounter{table}{0}
        \renewcommand{\thetable}{S\arabic{table}}%
        \setcounter{figure}{0}
        \renewcommand{\thefigure}{S\arabic{figure}}%
        \setcounter{equation}{0}
        \renewcommand{\theequation}{S\arabic{equation}}%
        \renewcommand{\figurename}{Supplementary Figure}
        \renewcommand{\tablename}{Supplementary Table}
        \setcounter{page}{1}
}

\beginsupplement
\onecolumngrid
\section*{Supplementary Material}
\subsection*{Supplementary Note: Force-extension curves of worm-like DNA}

\begin{figure}[h!]
    \centering
    \includegraphics[width=0.7\textwidth]{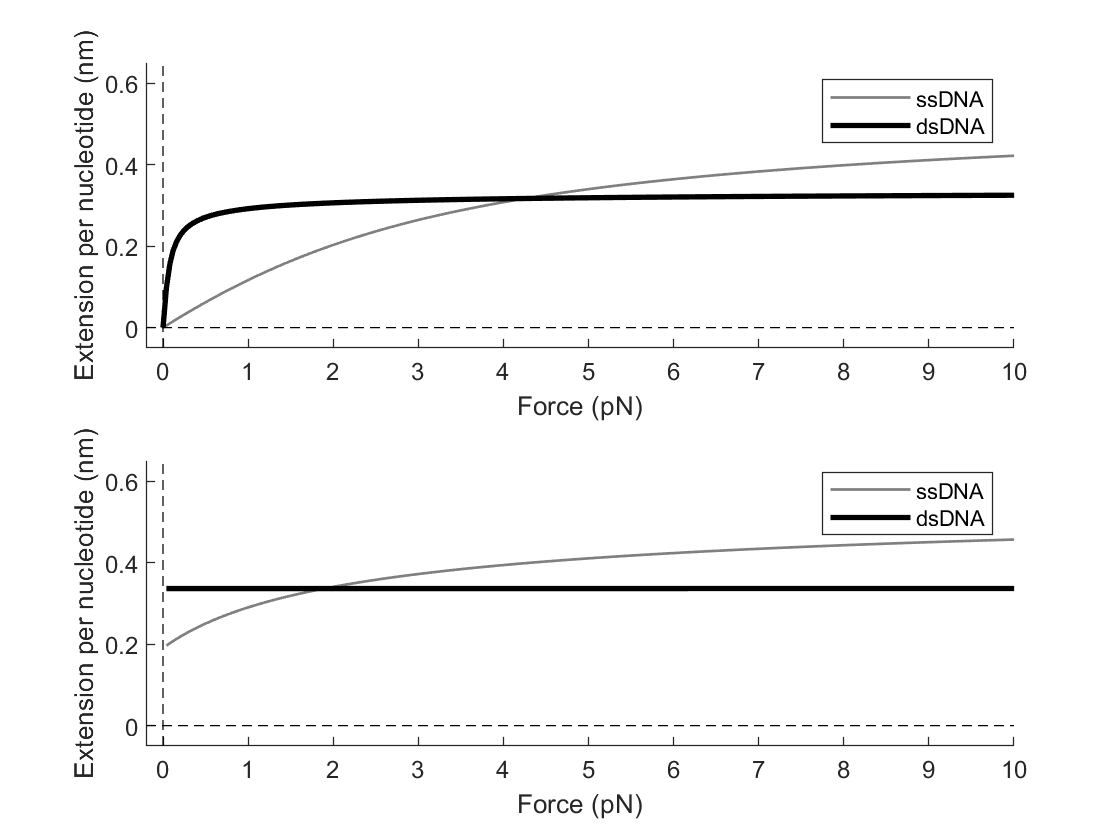}
    \caption{}
    \label{sfig:FECs}
\end{figure}

In the limit of $L\gg A$, Marko and Siggia derived an interpolation formula (MS formula) for the relationship between force ($f$) and extension ($x$) of a worm-like chain (WLC) \cite{marko1995stretching}:
\begin{equation}
    \frac{fA}{k_\mathrm{B}T}=\frac{x}{L}+\frac{1}{4\left(1-x/L\right)^2}-\frac{1}{4}
    \label{seq:MS}
\end{equation}
where $A$ is the persistence length, and $L$ is the contour length. It is convenient to define the contour length per nucleotide, $b=L/N$. The accuracy of this formula can be increased with additional terms \cite{bouchiat1999estimating}. Whitley et al. \cite{whitley2017elasticity} and Guo et al. \cite{guo2019understanding} modeled ssDNA and dsDNA as WLCs and also attempted modeling the transition state as a chimeric DNA of ssDNA and dsDNA or a WLC with its own unique $A$ and $L$. Using 53 nm and 0.34 nm for $A$ and $b$ of dsDNA, and 1.32 nm and 0.6 nm for $A$ and $b$ of ssDNA in Equation~\ref{seq:MS} and inverting it, we can obtain $x$ as a function of $f$ (top, Supplementary Figure~\ref{sfig:FECs}).
The extensions of ssDNA and dsDNA are predicted to cross over at $f \approx 4.3$.
A different formula that is more correct for short WLC is derived by Keller et al. \cite{keller2003relating} and Hori et al. \cite{hori2007stretching}. In this formula, $x$ is expressed as a function of $f$:
\begin{equation}
    x=L-\frac{k_\mathrm{B}T}{2f}\left( L\sqrt{\frac{f}{A k_\mathrm{B}T}}\coth\left(L\sqrt{\frac{f}{A k_\mathrm{B}T}}\right)-1\right).
    \label{seq:KH}
\end{equation}
Force-extension curves of ssDNA and dsDNA obtained from this formula are shown at the bottom of Supplementary Figure~\ref{sfig:FECs}. The crossover force is $\sim1.8$ pN, markedly lower than predicted by the MS formula.

\subsection{The role of stacking in hybridization}

\begin{figure}[h!]
    \centering
    \includegraphics[width=0.5\textwidth]{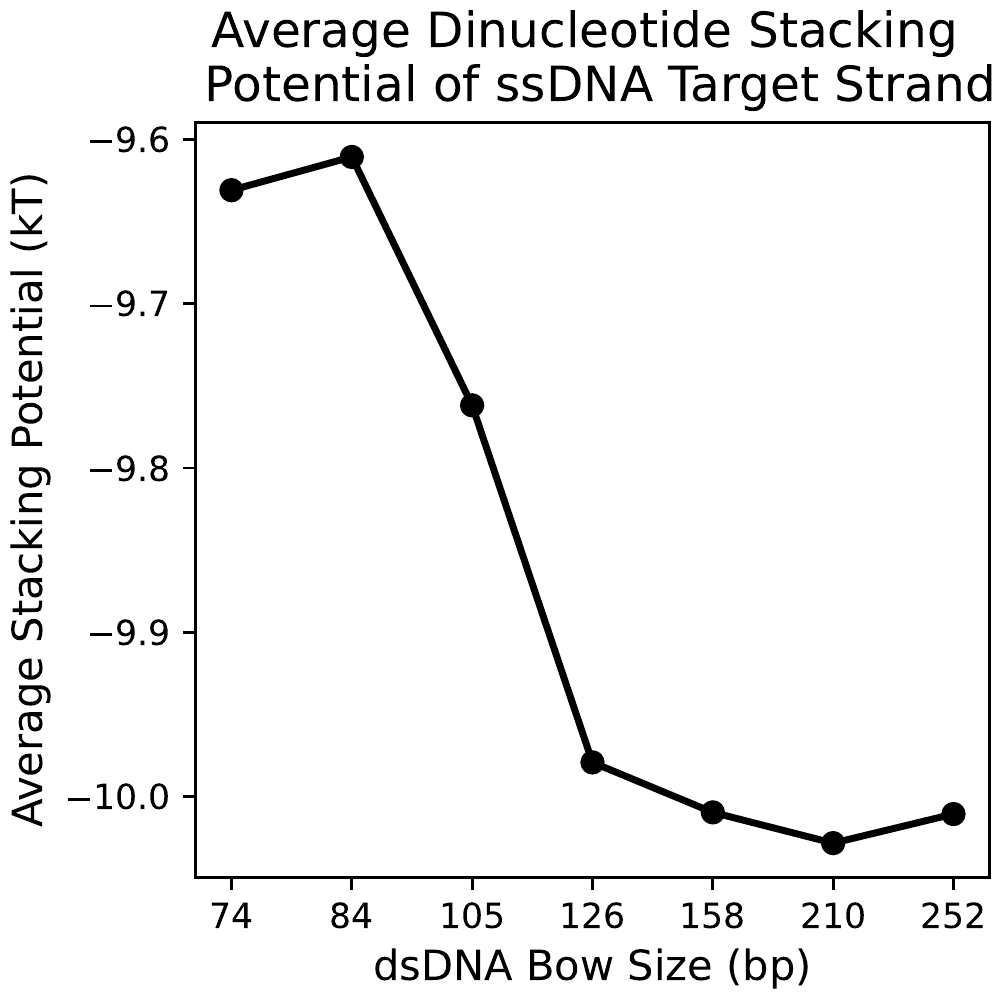}
    \caption{Average dinucleotide stacking potential of the ssDNA target sequence T2 for all DNA bow sizes. Energy values for each bow size are averaged over all saved configurations $7.5\times10^4$ and over all dinucleotide pairs in the target region. Note that stacking interactions between the terminal base pair of the dsDNA bow and the adjacent unpaired base in the ssDNA target are included in the average.}
    \label{sfig:bow_stacking_energy}
\end{figure}

Previous studies have hypothesized that weak DNA tension promotes binding by ordering ssDNA into ``prehelical" structures 
\cite{dupuis2013single,holbrook1999enthalpy}. However, when comparing the average stacking interactions of the ssDNA target strand for all DNA bow sizes, we observe that stacking interactions are weakest in our smallest DNA bows, for which our experimental results show higher hybridization rates (Supplementary Figure \ref{sfig:bow_stacking_energy}).
Moreover, our kinetics simulation indicates that stacking itself may have a negative impact on initial pairing between two complementary strands (Figure \ref{fig:binding_unbinding_heatmaps}). These two results are counterintuitive given that stacking stabilizes dsDNA. However, for bases on opposite strands to pair, they need rotational freedom, which would be restricted if base stacking is present. Therefore, base stacking seems to play conflicting roles in both hindering base pair formation prior to strand docking as well as stabilizing base pairs once they are formed.

\subsection*{Supplementary Information: oxDNA2 simulation parameters}
For all oxDNA2 simulations, the buffer conditions were specified to be identical to our experiments (\SI{100}{\milli\molar} salt concentration and \SI{22}{\celsius}). To prevent non-representative initial states, all simulations were equilibrated for 50000 time steps before configurations were saved into output trajectories \cite{sengar2021primer}. All simulations used an Andersen-like thermostat \cite{russo2009reversible}, where the molecular system was propagated according to Newton's equations for $N_\mathrm{Newt}$ time steps using Verlet integration; afterward, the system was assigned new linear and angular velocities drawn from a Maxwell-Boltzmann distribution such that the resulting diffusion coefficient was equal to a specified value $D$. For all DNA bow simulations and force-extension simulations, $N_\mathrm{Newt} = 103$ and $D=2.5$; for all FFS simulations, $N_\mathrm{Newt} = 51$ and $D=1.25$.

\subsection*{Supplementary Method: Estimating the end-to-end distance radial probability distribution of DNA bows}

To estimate the tensile force $f$ exerted by each DNA bow size (Equation \ref{eq:tensile_force}), we used the following interpolation formula to estimate the radial probability distribution $P(x)$ of a wormlike chain

\begin{equation} \label{eq:radial_distribution}
\begin{aligned}
P(x') = 4\pi x'^{\; 2} \cdot J_{SYD} \cdot \Bigg(\frac{1-cx'^{2}}{1-x'^2}\Bigg)^{5/2}\exp\Bigg(\frac{\sum_{i=-1}^{0}\sum_{j=1}^{3}c_{i,j}\kappa^ix'^{2j}}{1-x'^2}\Bigg) \\ \times\exp\Bigg(\frac{-d\kappa ab(1+b)x'^2}{1-b^2x'^2}\Bigg)I_0\Bigg(-\frac{d\kappa a b (1+b)x'^2}{1-b^2x'^2}\Bigg),
\end{aligned}
\end{equation}
where
\begin{equation*}
    a  = 14.054, \quad b = 0.473, \quad (c_{i,j})_{i,j} = \begin{pmatrix} 
    -3/4 & 23/64 & -7/64 \\ 
    -1/2 & 17/16 & -9/16
    \end{pmatrix}.
\end{equation*}
This formula accurately models $P(x)$ for a large range of stiffness values ($\kappa = A / L$, where $A$ and $L$ are the persistence and contour lengths of the dsDNA elastic arc, respectively) as well as a wide range of normalized end-to-end distance values \cite{becker2010radial} ($x' = x/L$). For this calculation, we assumed the values $A=\SI{53}{\nano\meter}$ and $b = \SI{0.34}{\nano\meter}$, where $b$ is the contour length per nucleotide $b=L/N$  Therefore, Equation \ref{eq:radial_distribution} can be used with Equation \ref{eq:tensile_force} to estimate the force exerted by all bow sizes, whose stiffnesses range from $\kappa=0.6$ to $\kappa$=2.1, and whose end-to-end distance values range from $x=0.06$ to $x=0.21$.

\subsection*{Supplementary Method: Simulating unbinding and binding reactions with forward flux sampling (FFS)}

For all FFS simulations, the center of mass of each of the terminal bases on the \SI{17}{\nt} target molecule T1 were separated by a fixed extension value $x$ using two strong harmonic traps with force constant k = \SI{570.9}{\pico\newton/\nano\meter} (10 simulation units in oxDNA). $x$ was fixed at $\SI{5.5}{\nano\meter}$ for unbinding reactions and $\SI{5.1}{\nano\meter}$ for binding reactions, which are equivalent to the average extension values observed for our largest DNA bow in its bound and unbound state respectively (Tables \ref{table:e2e_distances}). All FFS simulations were performed using a \SI{9.1}{\femto\second} time step. 

The unbinding FFS simulation was separated into 9 interfaces (Supplementary Table \ref{table:unbinding_order_parameters}. Each interface $\lambda_Q$ corresponded to a change in the number of remaining base pairs, decreasing from 8 remaining base pairs to 0 base pairs. Base pairing was defined as when any two complementary bases had a hydrogen bond potential energy less than -0.1 simulation units ($\SI{-0.6}{\kilo \calorie/\mole}$, or about $1\ k_B T$ at \SI{22}{\celsius}). The initial flux was calculated by running a brute force trajectory of the molecule in state A  (\SI{9}{\bp}) and observing the rate of forward crossings across the first interface $\lambda_0$ (\SI{8}{bp}) according to
\begin{equation}\label{eq:initial_flux}
    \Phi_{A,0} = \frac{N_0}{T}, 
\end{equation}
where $N_0$ is the number of crossings and T is the total time duration of trajectories where A was more recently visited than B. Using the configurations of successful crossings saved during initial flux simulation, the transition probability $P(\lambda_1|\lambda_0)$ of melting the next base pair (\SIrange{8}{7}{\bp}) was calculated according to 
\begin{equation}\label{eq:melt_prob}
    P(\lambda_1|\lambda_0) = \frac{N_1}{M_0},
\end{equation}
where $M_0$ is the number of trial trajectories started at $\lambda_0$ and $N_1$ is the number of trajectories that successfully reach $\lambda_1$. Note that trajectories are halted and marked as failures upon reaching state A. Each trial trajectory is started from a configuration randomly selected from the $N_0$ configurations saved during the initial flux simulation. In the following steps, we implemented a variant of FFS known as ``pruning", which eliminates a large fraction of backward-moving trajectories and re-weights the surviving trials \cite{allen2006simulating}. In these steps, trajectories that revert backward to interface $\lambda_{Q-2}$ after starting at $\lambda_{Q-1}$ were pruned with probability $p=0.75$. To correct for this, the transition probability $P(\lambda_Q|\lambda_{Q-1})$ of melting additional base pairs was calculated according to
\begin{equation}\label{eq:prune_melt_prob}
    P(\lambda_Q|\lambda_Q-1) = \frac{N_Q-N_Q^*+N_Q^*/(1-p)}{M_{Q-1}},
\end{equation}
where $M_{Q-1}$ is the number of trial trajectories started at $\lambda_{Q-1}$, $N_Q$ is the total number of trajectories that successfully reach $\lambda_Q$, and $N_Q^*$ is the number of trajectories that revert to $Q-2$, survive pruning, and ultimately reach $\lambda_Q$. The initial flux of crossing $\lambda_{0}$, as well as the probabilities of crossing successive interfaces, are tabulated in \ref{table:ffs_results}.

The binding FFS simulation was separated into 11 interfaces. Similar to unbinding, the initial flux as well as the subsequent melting probabilities were calculated using Equations \ref{eq:initial_flux}--\ref{eq:melt_prob}.  As before, pruning was implemented for interfaces after $\lambda_0$. Interfaces for the first two ``strand approach" steps were defined with a distance order parameter $d$, where $d$ was defined as the minimum separation between any two complementary bases on the probe and target segment. Similar to unbinding, interfaces for the remaining 9 steps corresponded to a change in the number of paired bases $n$ in the partial duplex, starting at \SI{1}{\bp} and ending at \SI{9}{\bp} (Supplementary Table \ref{table:binding_order_parameters}). Similar to unbinding, base pairs were defined as when any two complementary bases had a hydrogen bond potential energy of less than -0.1 simulation units ($\SI{-0.6}{\kilo \calorie/\mole}$). The initial flux of crossing $\lambda_{0}$, as well as the probabilities of crossing each successive interface ($\lambda_Q$), are tabulated in \ref{table:ffs_results}.

\begin{figure}
    \centering
    \includegraphics[width=\textwidth]{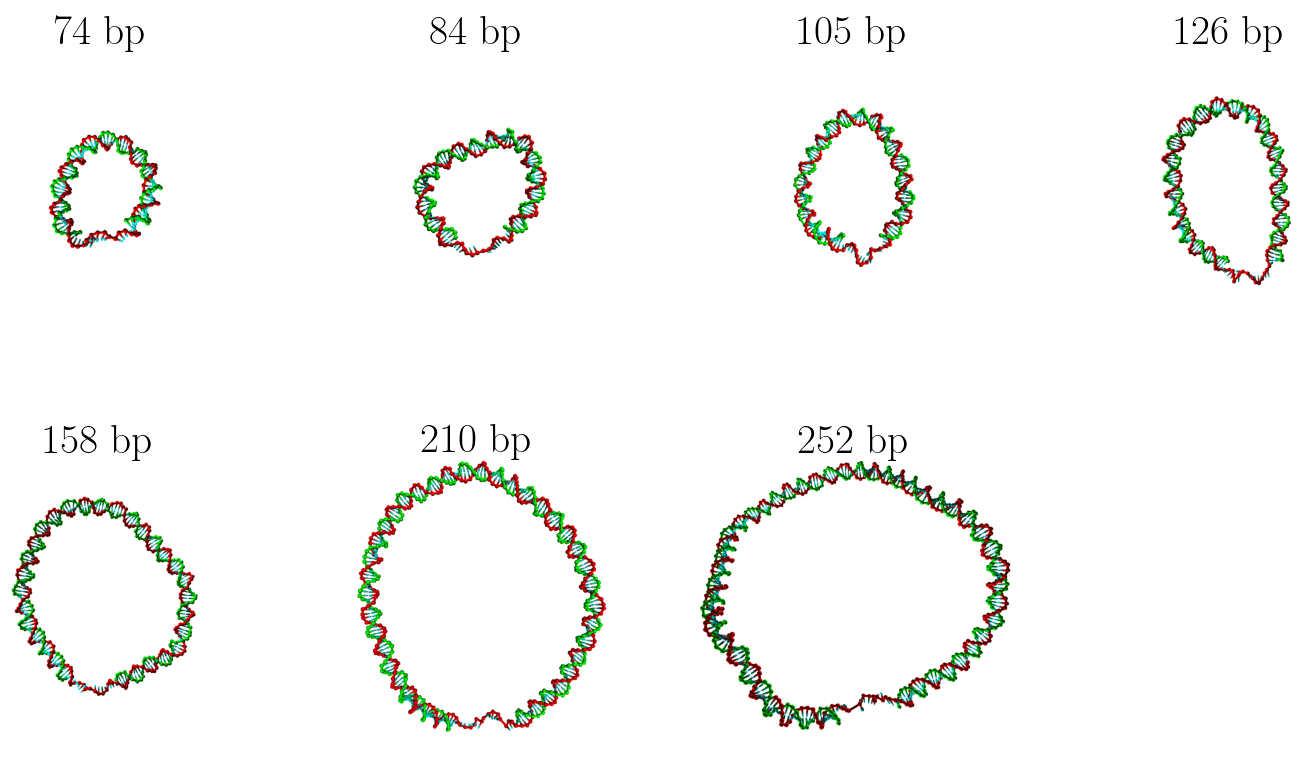}
    \caption{Sample oxDNA2 configurations of all experimentally measured DNA bow sizes. Each configuration depicts the DNA bow in its unbound state. The label above each molecule specifies the length of the dsDNA bow arc (in base pairs). All constructs feature a \SI{15}{\nt} ssDNA strand containing a \SI{9}{\nt} region targeted by an 8-9 \SI{}{\nt} probe.}
    \label{sfig:configs_dna_bow_array}
\end{figure}

\begin{figure}
    \centering
    \adjincludegraphics[width = \textwidth]{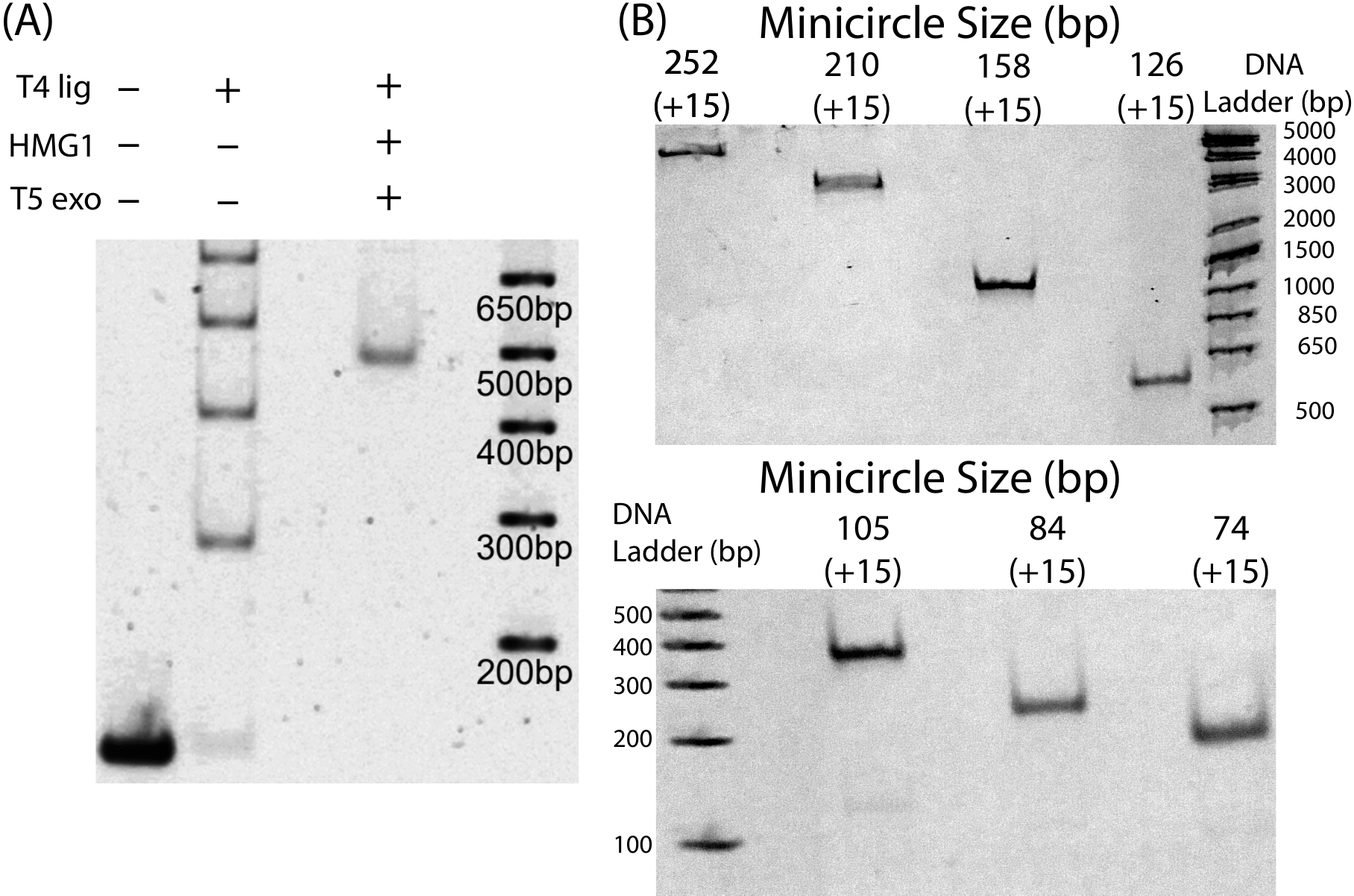}
    \caption{\textbf{(A)} Linear DNA molecules with phosphorylated 5’ ends were bent with bending protein HMG1 and self-ligated. T5 exonuclease was then added to digest unwanted polymer fragments. \textbf{(B)} The remaining circular DNA was then purified with ethanol precipitation and nicked on the unmodified strand with Nb.BbvCI. Nicked minicircle bands were analyzed and extracted using polyacrylamide gel electrophoresis (6\%, 29:1 acrylamide to bis-acrylamide in 0.5x TBE buffer). Note that the total minicircle size includes the \SI{15}{\bp} target strand segment, which is not included in the bow arc length (\SIrange{74}{252}{\bp}. After each circle size was inspected via PAGE, DNA minicircles were extracted overnight via ``crush-and-soak" and concentrated with ethanol precipitation.}
    \label{sfig:mc_gels}
\end{figure}

\begin{figure}
    \centering
    \includegraphics[width=0.75\textwidth]{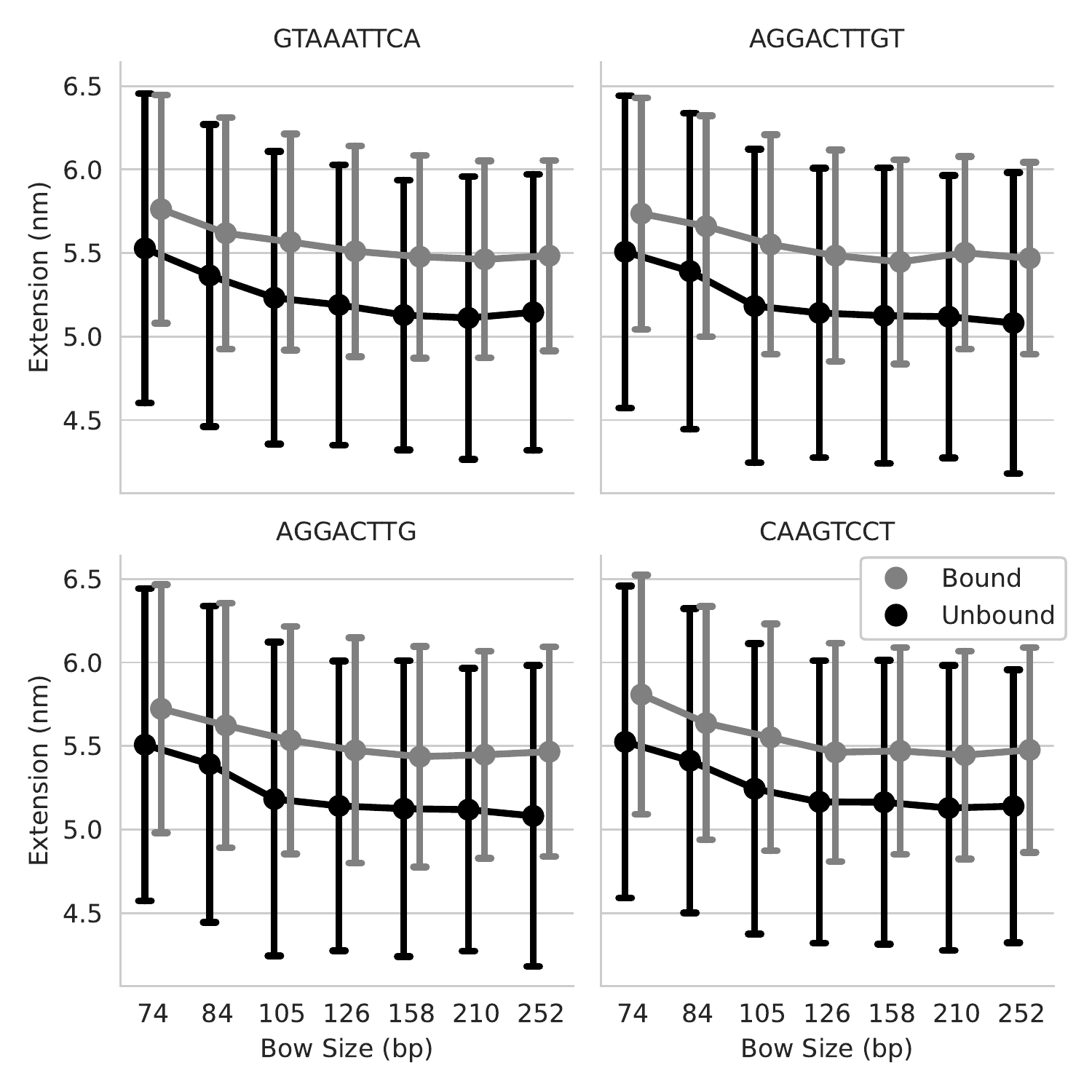}
    \caption{ Mean ($\overline{x}$) and standard deviation values ($\sigma(x)$) of the end-to-end distance $x$ for each DNA bow size, in both the probe-bound and probe-unbound states. Each MD simulation was performed for $7.5\times10^7$ steps with $dt = \SI{15.2}{\femto\second}$, totaling $t = \SI{1.14}{\micro\second}$. Extension values were measured every 1000 steps, collecting $n=7.5\times10^4$ values in total.}
    \label{sfig:dna_bow_distributions}
\end{figure}

\begin{figure}
    \centering
    \includegraphics{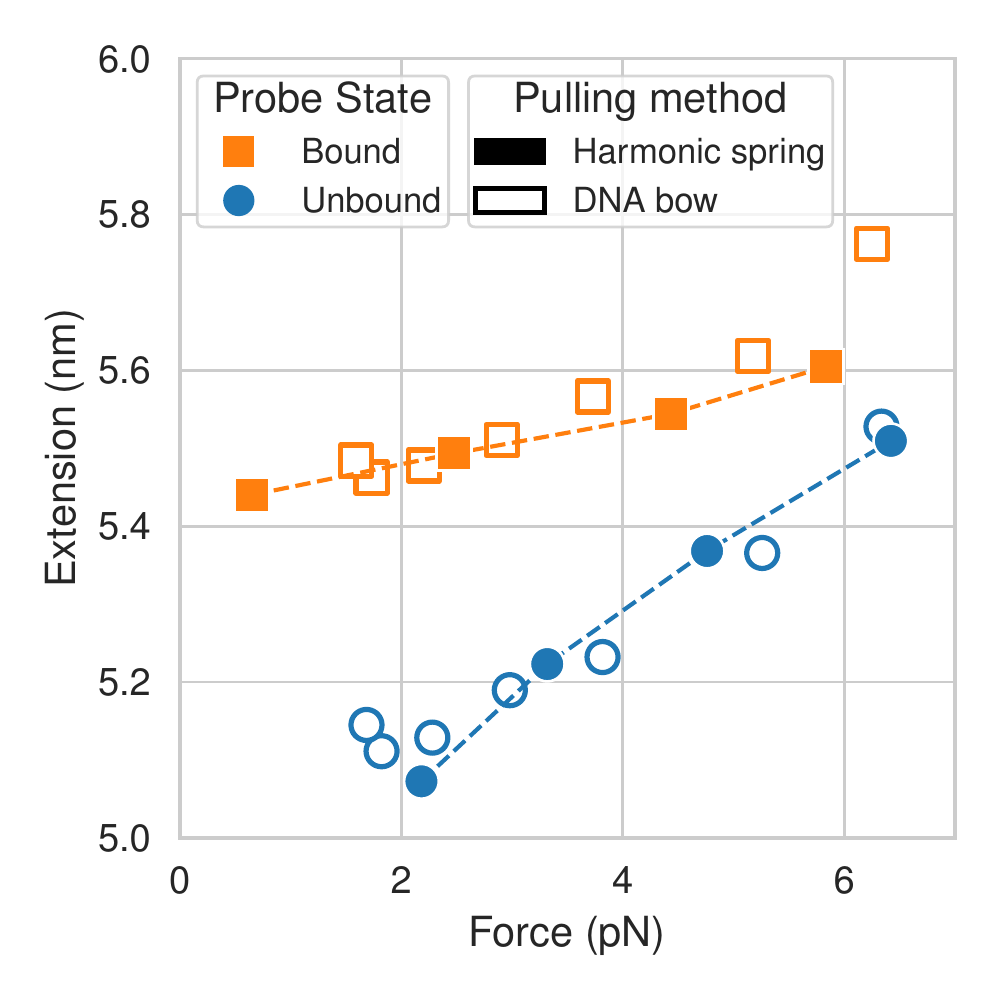}
    \caption{Force-extension behavior of the \SI{17}{\nt} target sequence T1 in both its probe-bound and probe-unbound states, extended by either a harmonic spring (filled markers) or a DNA bow (open markers). For the harmonic spring pulling method, we used the ``mutual trap" external force tool provided with oxDNA to connect the terminal bases of the target with a weak spring.}
    \label{sfig:spring_vs_bow}
\end{figure}

\begin{figure}
    \centering
    \includegraphics[width=0.8\textwidth]{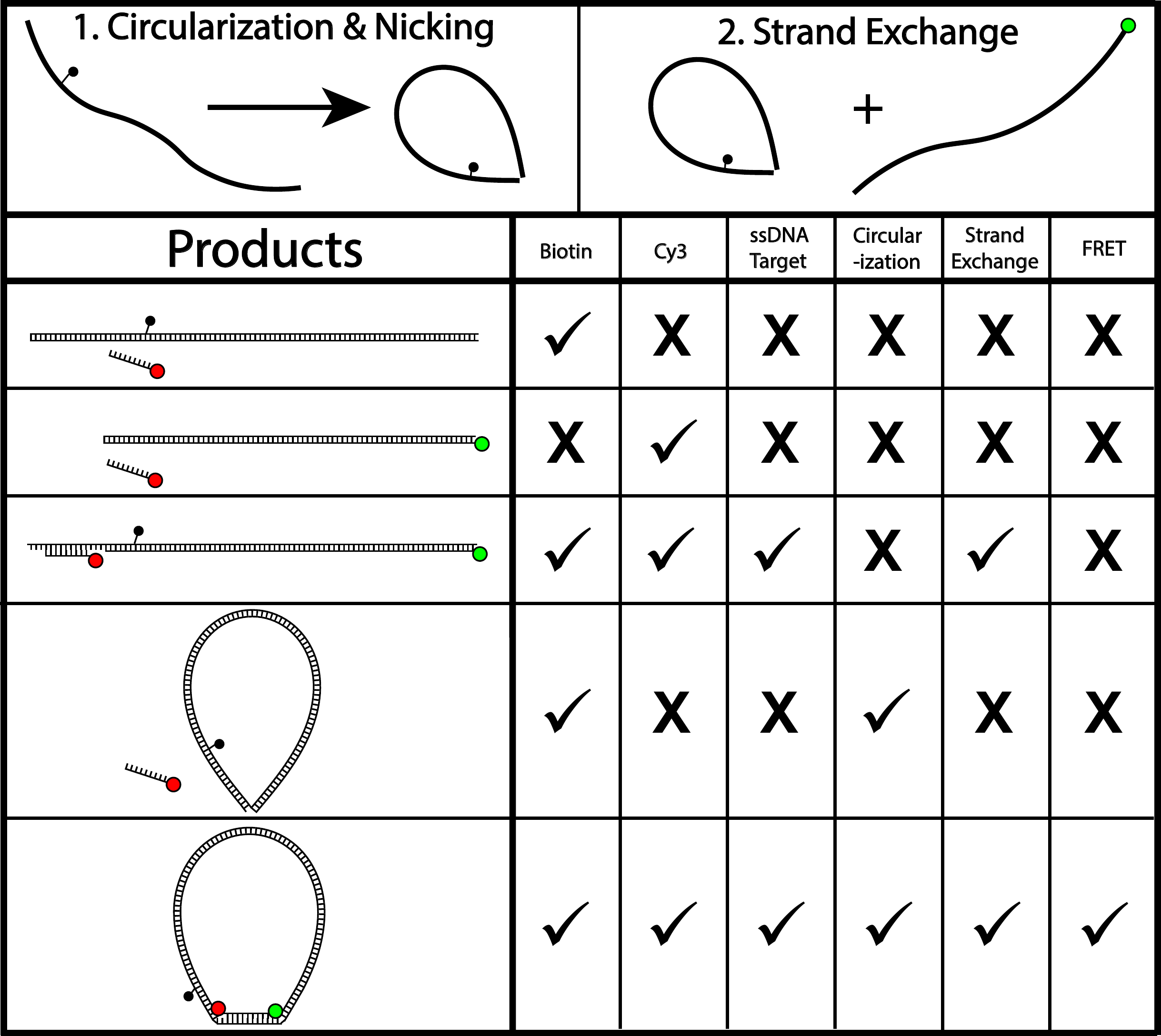}
    \caption{Possible products created during bow construction. Nicked circular products were purified and mixed with Cy3-labeled linear molecules at a 1:4 ratio. The unmodified nicked strand is replaced via a strand exchange reaction which consists of heating the mixture to \SI{95}{\celsius} and cooling to \SI{4}{\celsius} gradually. By design, only the desired product (bottom row) is capable of generating a FRET signal. Incorrect purification of nicked circular strands (step one) would yield linear products during strand exchange; among these products, the target is either too far from the Cy3 dye to generate a FRET signal upon probe binding, or the target is absent entirely. Circular molecules that do not replace the unmodified strand during the strand exchange reaction (step two) are not donor-labeled, nor do they have an exposed acceptor-probe target, and therefore also cannot generate a FRET signal.}
    \label{sfig:dna_bow_table}
\end{figure}

\begin{figure}
    \centering
    \includegraphics[width=0.7\textwidth]{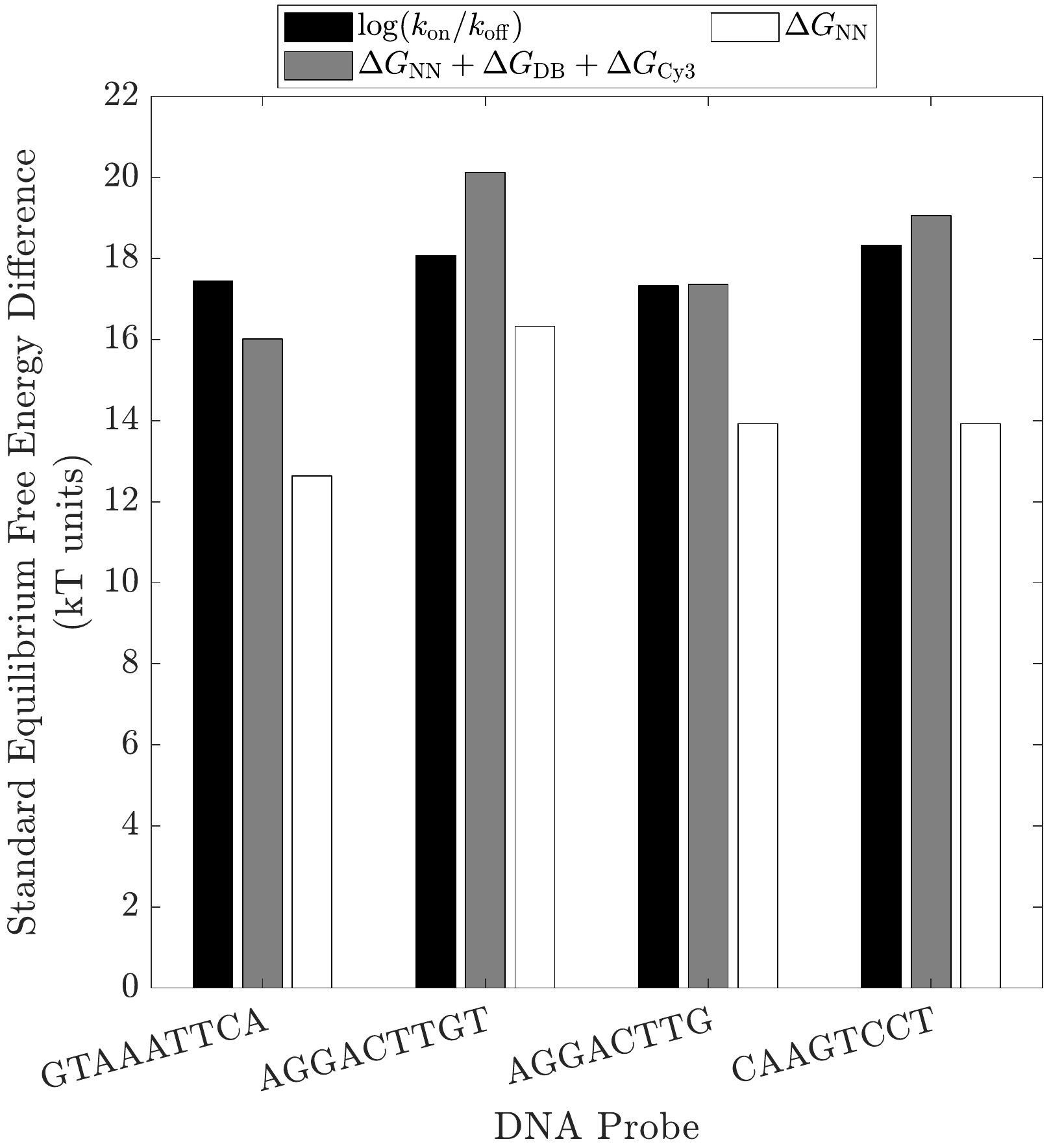}
    \caption{A comparison of the measured standard equilibrium free energy difference values of all DNA probes to their corresponding nearest-neighbor predictions. The free energy difference, $\Delta G = \log(k_\mathrm{on}/k_\mathrm{off})$, was calculated using the average $k_\mathrm{off}$ and $k_\mathrm{on}$ values observed for the \SI{252}{\bp} DNA bow (black bars). The predicted free energy difference of a freely diffusing \SIrange{8}{9}{\bp} duplex, $\Delta G_\mathrm{NN}$, was then calculated using published nearest-neighbor thermodynamic parameters (white bars) \cite{santalucia1996improved}. We then modified this estimate to correct for our experimental conditions by adding the energy contribution of dangling base stacking interactions $\Delta G_\mathrm{DB}$ \cite{bommarito2000thermodynamic} as well as the energy contribution of a Cy3 dye attachment $\Delta G_\mathrm{Cy3}$ (gray bars) \cite{moreira2015cy3}. Both estimates were calculated at a temperature \SI{22}{\celsius} to match our experiment. The resulting sum was corrected for the monovalent cation concentration of our buffer ($\mathrm{[Mono^+] = [Na^+] + [Tris^+] = \SI{150}{\milli\molar}}$) using the calibration formula published by SantaLucia Jr. Note that this estimate assumes that half of Tris molecules are protonated \cite{owczarzy2008predicting}.}
    \label{sfig:NN_barplot}
\end{figure}

\begin{figure}
    \centering
    \includegraphics[width=\textwidth]{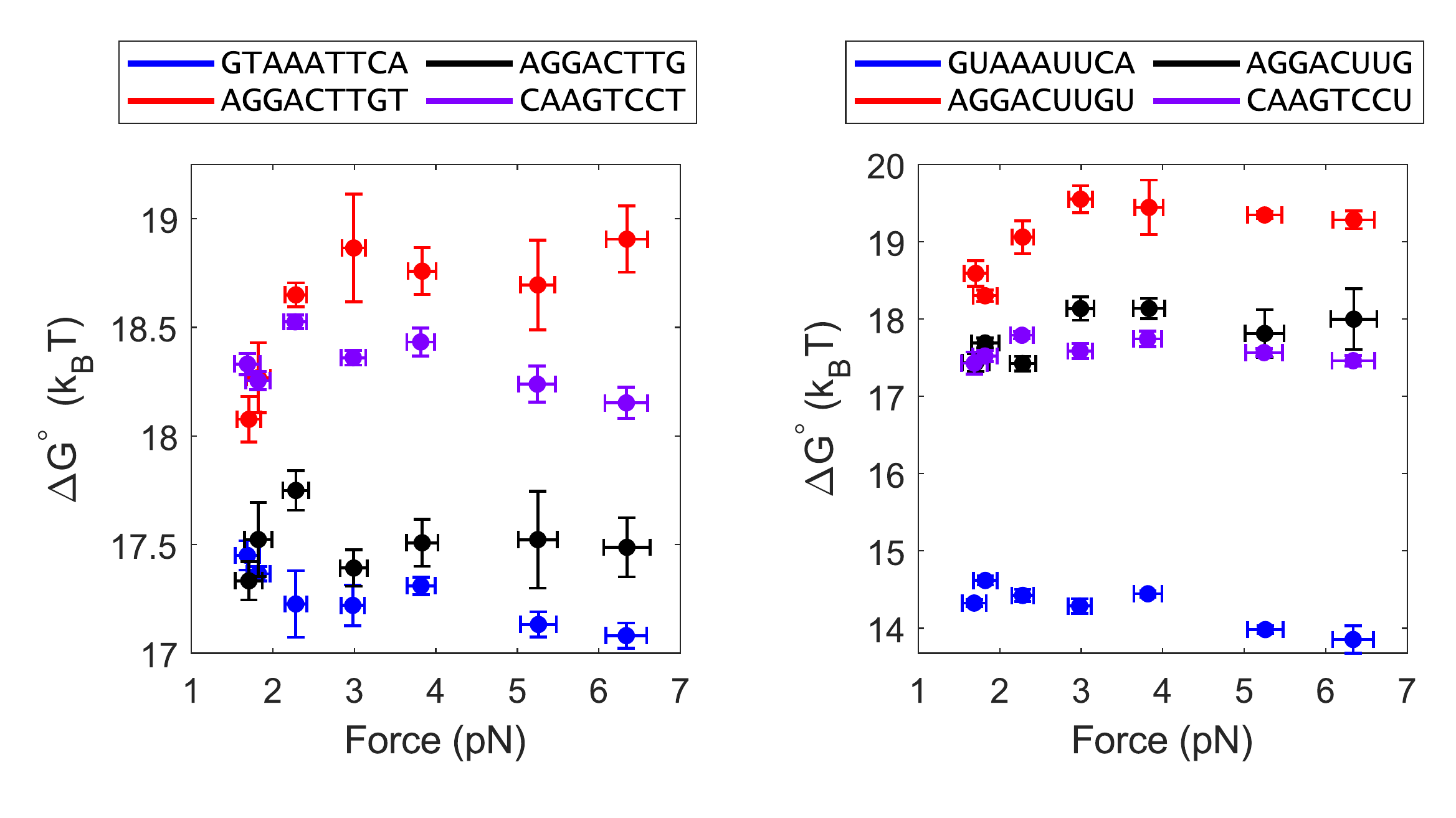}
    \caption{Force dependence of the equilibrium free energy difference $\Delta G = \log(k_\mathrm{on}/k_\mathrm{off})$ between the bound and unbound states. Here, $\Delta G$ is defined as the additional free energy of the unbound state relative to the bound state. The force exerted by each bow was calculated with Equation \ref{eq:tensile_force}, using the mean extension $\overline{x}$ of the bow's end-to-end distance distribution. Vertical error bars represent the standard error of the mean; horizontal error bars were calculated using $\frac{\partial f(x)}{\partial x}\Big|_{\overline{x}} \cdot \sigma(x)$, where $\overline{x}$ and $\sigma(x)$ are the mean and standard deviation of the bow's end-to-end distance distribution.}
    \label{sfig:G_vs_f}
\end{figure}

\begin{figure}
    \centering
    \includegraphics[width=\textwidth]{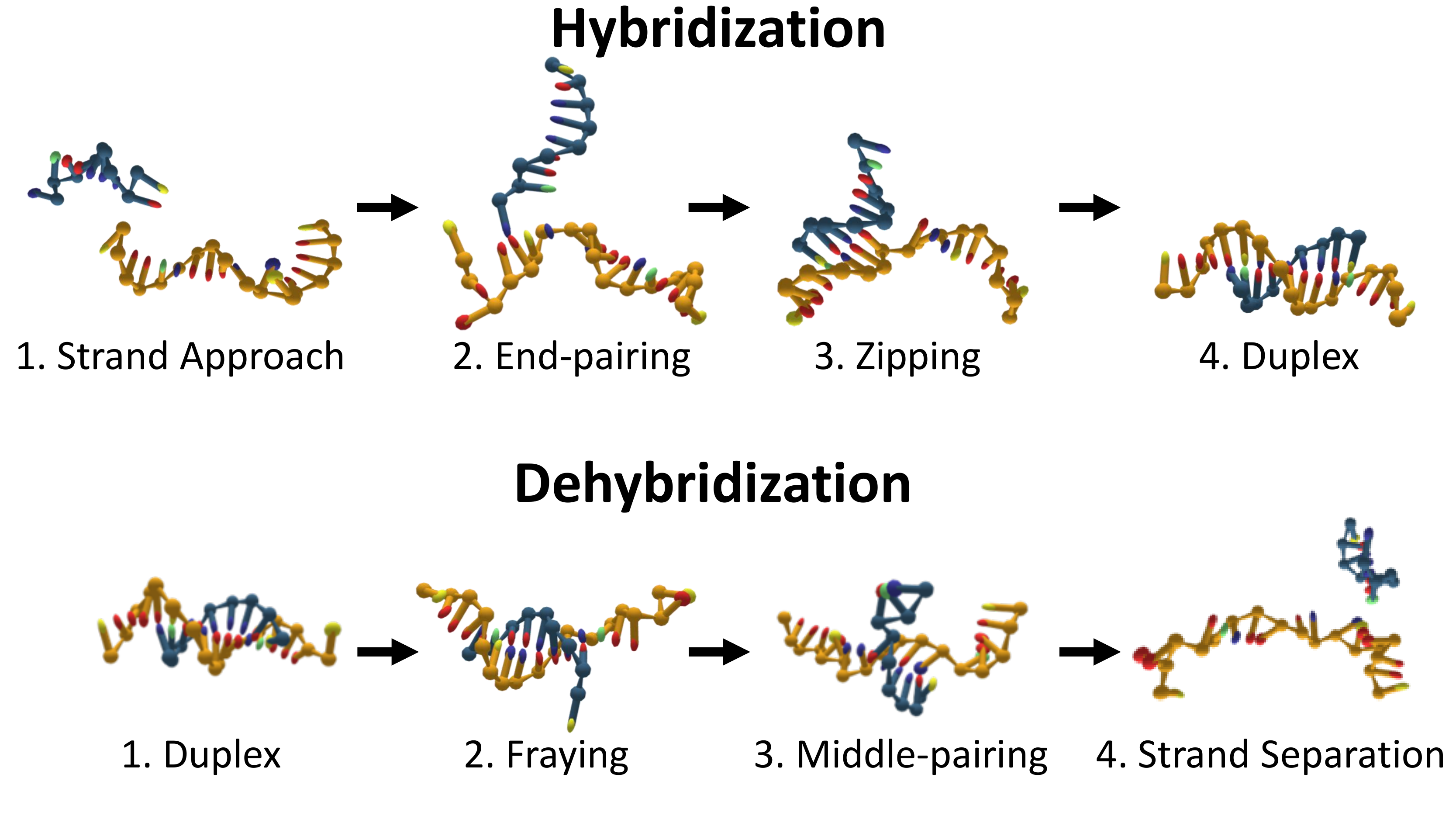}
    \caption{Flowchart of hybridization and dehybridization reactions. During a typical binding transition, two strands approach one another and form an initial base pair at their terminal ends; afterwards, the two strands ``zip" together in a linear fashion. During a typical unbinding transition, the base pairs at the ends of the two strands fray and separate, continuing inward until one last base pair remains near the center; after this final middle-pair melts, the strand separates.}
    \label{sfig:DNA_reaction_flowchart}
\end{figure}




\clearpage
\newpage



\clearpage
\LTcapwidth=\textwidth
 \renewcommand{\arraystretch}{.75}
 
\begin{longtable}[H]{|p{.15\linewidth}| >{\setlength{\baselineskip}{.75\baselineskip}}p{0.8\linewidth}|}
 
    \hline
    \multicolumn{2}{|c|}{\textbf{DNA bow arc duplex segments ($5^\prime$ to $3^\prime$)}}\\
    \hline
    74 bp &  \texttt{GACTCCCCACTCGTCGTACGAGGTCGCACACGCCCCACACCCAGACCTCCCTGCCCTGGTACCTC AGCACTGAG}\\
    \hline
    84 bp &  \texttt{GACTCCCCACTCGTCGTACGCAACGAGGTCGCACACGCCCCACACCCAGACCTCCCTGCGAGCGC CTGGTACCTCAGCACTGAG}\\
    \hline
    105 bp & \texttt{GACTCCCCACTCGTCGTACGATCGCCATGGCAACGAGGTCGCACACGCCCCACACCCAGACCTCC CTGCGAGCGGGCATGGGTACCCTGGTACCTCAGCACTGAG}\\
    \hline
    126 bp & \texttt{GACTCCCCACTCGTCGTACCACCCACGCGCGATCGCCATGGCAACGAGGTCGCACACGCCCCACA CCCAGACCTCCCTGCGAGCGGGCATGGGTACAATGTCCCCGCCTGGTACCTCAGCACTGAG}\\
    \hline
    158 bp & \texttt{GACTCCCCACTCGTCGTACGTTTGGGGAAAGACCACACCCACGCGCGATCGCCATGGCAACGAGG TCGCACACGCCCCACACCCAGACCTCCCTGCGAGCGGGCATGGGTACAATGTCCCCGTTGCCACA GAGACCACCCTGGTACCTCAGCACTGAG}\\
    \hline
    210 bp & \texttt{GACTCCCCACTCGTCGTACTGCGAAATCCGGAGCAACGGGCAACCGTTTGGGGAAAGACCACACC CACGCGCGATCGCCATGGCAACGAGGTCGCACACGCCCCACACCCAGACCTCCCTGCGAGCGGGC ATGGGTACAATGTCCCCGTTGCCACAGAGACCACTTCGTAGCACAGCGCAGAGCGTAGCGCCTGG TACCTCAGCACTGAG}\\
    \hline
    252 bp & \texttt{GACTCCCCACTCGTCGTACTTTTTGTTTACGCGACAACTATGCGAAATCCGGAGCAACGGGCAAC CGTTTGGGGAAAGACCACACCCACGCGCGATCGCCATGGCAACGAGGTCGCACACGCCCCACACC CAGACCTCCCTGCGAGCGGGCATGGGTACAATGTCCCCGTTGCCACAGAGACCACTTCGTAGCAC AGCGCAGAGCGTAGCGTGTTGTTGCTGCTGACAAAAGCCTGGTACCTCAGCACTGAG}\\
    \hline
    \multicolumn{2}{|c|}{\textbf{Primers for making DNA force assay duplex segments ($5^\prime$ to $3^\prime$)}}\\
    \hline
    & \texttt{{} {} {} {} {} {} {} {} {} {} {} {} {} {} {} {} {} {} {} $\downarrow$ 20 nt }\\
    \hline
    74 Forward & \texttt{GACTCCCCACTCGTCGTACGAGGTCGCACACGCC}\\
    \hline
    84 Forward & \texttt{GACTCCCCACTCGTCGTACGCAACGAGGTCGCACAC}\\
    \hline
    105 Forward & \texttt{GACTCCCCACTCGTCGTACGATCGCCATGGCAACG}\\
    \hline
    126 Forward & \texttt{GACTCCCCACTCGTCGTACCACCCACGCGCGAT}\\
    \hline
    158 Forward & \texttt{GACTCCCCACTCGTCGTACGTTTGGGGAAAGACCACAC}\\
    \hline
    210 Forward & \texttt{GACTCCCCACTCGTCGTACTGCGAAATCCGGAGCA}\\
    \hline
    252 Forward & \texttt{GACTCCCCACTCGTCGTACTTTTTGTTTACGCGACAACTATG} \\
    \hline
    
    74 Reverse & \texttt{CTCAGTGCTGAGGTACCAGGGCAGGGAGGTCTGGGTG}\\
    \hline
    84 Reverse & \texttt{CTCAGTGCTGAGGTACCAGGCGCTCGCAGGGAGGT}\\
    \hline
    105 Reverse & \texttt{CTCAGTGCTGAGGTACCAGGGTACCCATGCCCGCTC}\\
    \hline
    126 Reverse & \texttt{CTCAGTGCTGAGGTACCAGGCGGGGACATTGTACCCATG}\\
    \hline
    158 Reverse & \texttt{CTCAGTGCTGAGGTACCAGGGTGGTCTCTGTGGCAACG}\\
    \hline
    210 Reverse & \texttt{CTCAGTGCTGAGGTACCAGGCGCTACGCTCTGCGCT}\\
    \hline
    252 Reverse & \texttt{CTCAGTGCTGAGGTACCAGGCTTTTGTCAGCAGCAACAACA}\\
    \hline
    \multicolumn{2}{|c|}{\textbf{Primers for making circular molecules, target segment underlined ($5^\prime$ to $3^\prime$)}}\\
    \hline
    T1 & \texttt{[Phos]TTT\underline{TGAATTTAC}TTTGACTCCCCAC[BiotindT]CGTCGTAC}\\
    \hline
    T2 \& T3 & \texttt{[Phos]TTT\underline{ACAAGTCCT}TTTGACTCCCCAC[BiotindT]CGTCGTAC}\\
    \hline
    T4 & \texttt{[Phos]TTT\underline{AGGACTTGT}TTTGACTCCCCAC[BiotindT]CGTCGTAC}\\
    \hline
    Reverse & \texttt{[Phos]CTCAGTGC|TGAGGTACCAGG}\\
    \hline
    \multicolumn{2}{|c|}{\textbf{Primers for making Cy3-labeled molecules for strand exchange ($5^\prime$ to $3^\prime$)}}\\
    \hline
    Forward & \texttt{GACTCCCCACTCGTCGTAC}\\
    \hline
    Reverse (Cy3) & \texttt{[Cy3]CTCAGTGCTGAGGTACCAGG}\\
    \hline
    \multicolumn{2}{|c|}{Cy5 acceptor probes ($5^\prime$ to $3^\prime$)}\\
    \hline
    \multicolumn{2}{|c|}{\textbf{DNA and RNA probes for smFRET experiments ($5^\prime$ to $3^\prime$)}}\\
    \hline
    P1-DNA & \texttt{[Cy5]GTAAATTCA}\\ \hline
    P1-RNA & \texttt{[Cy5]GUAAAUUCA}\\ \hline
    P2-DNA & \texttt{[Cy5]AGGACTTGT}\\ \hline
    P2-RNA & \texttt{[Cy5]AGGACUUGU}\\ \hline
    P3-DNA & \texttt{[Cy5]AGGACTTG}\\ \hline
    P3-RNA & \texttt{[Cy5]AGGACUUG}\\ \hline
    P4-DNA & \texttt{[Cy5]CAAGTCCT}\\ \hline
    P4-RNA & \texttt{[Cy5]CAAGUCCU}\\
    \hline
    \multicolumn{2}{|c|}{\textbf{FFS simulation sequences, ($5^\prime$ to $3^\prime$)}}\\
    \hline
    P1-DNA & \texttt{GTAAATTCA}\\ \hline
    T1 & \texttt{GTTT\underline{TGAATTTAC}TTTG}\\ \hline
    
    \caption{List of DNA sequences, PCR primers, and DNA/RNA probes. All bow arc duplex segments are sourced from yeast genomic DNA, and extended to include common adapter sequences on each end. Forward primers for making circular DNA include the \SI{15}{\nt} sequence containing the \SI{9}{\nt} ssDNA complementary target segment (underlined); the reverse primer for making circular DNA includes the nick site (marked with a vertical line ``\texttt{|}"). DNA and RNA probes were added to imaging buffer at \SI{20}{\nano\molar} during smFRET experiments to measure unbinding ($k_\mathrm{off}$) and binding rates ($k_\mathrm{on}$). Note that DNA target sequences used for FFS simulations include an additional nucleotide at each end, matching the letter of the terminal bases in the dsDNA portion of bow constructs.}
    \label{table:sequences}
\end{longtable}
\clearpage

\begin{table}
\renewcommand{\arraystretch}{.6} 
\centering
\begin{tabular}{|c|c|c|c|c|c|c|c|c|}
    \hline
    \multicolumn{8}{|c|}{End-to-end extension (nm), $\overline{x}\pm \sigma(x)$}\\
    \hline
    \multicolumn{8}{|c|}{Unbound state}\\\hline
    Target &\SI{74}{\bp}&\SI{84}{\bp}&\SI{105}{\bp}&\SI{126}{\bp}&\SI{158}{\bp}&\SI{210}{\bp}&\SI{252}{\bp}\\ \hline
    1      &$5.5\pm0.9$ &$5.4\pm0.9$ &$5.2\pm0.9$  &$5.2\pm0.8$  &$5.1\pm0.8$  &$5.1\pm0.8$  &$5.1\pm0.8$  \\ \hline
    2 \& 3 &$5.5\pm0.9$ &$5.4\pm0.9$ &$5.2\pm0.9$  &$5.1\pm0.9$  &$5.1\pm0.9$  &$5.1\pm0.8$  &$5.1\pm0.9$  \\ \hline
    4      &$5.5\pm0.9$ &$5.4\pm0.9$ &$5.2\pm0.9$  &$5.2\pm0.8$  &$5.2\pm0.8$  &$5.1\pm0.9$  &$5.1\pm0.8$  \\ \hline
    \multicolumn{8}{|c|}{Bound state}\\
    \hline
     Target &\SI{74}{\bp}&\SI{84}{\bp}&\SI{105}{\bp}&\SI{126}{\bp}&\SI{158}{\bp}&\SI{210}{\bp}&\SI{252}{\bp} \\ \hline
     1      &$5.8\pm0.7$ &$5.6\pm0.7$ &$5.6\pm0.6$  &$5.5\pm0.6$  &$5.5\pm0.6$  &$5.5\pm0.6$  &$5.5\pm0.6$   \\ \hline
     2      &$5.7\pm0.7$ &$5.7\pm0.7$ &$5.6\pm0.7$  &$5.5\pm0.6$  &$5.4\pm0.6$  &$5.5\pm0.6$  &$5.5\pm0.6$   \\ \hline
     3      &$5.7\pm0.7$ &$5.6\pm0.7$ &$5.5\pm0.7$  &$5.5\pm0.7$  &$5.4\pm0.7$  &$5.4\pm0.6$  &$5.5\pm0.6$   \\ \hline
     4      &$5.8\pm0.7$ &$5.6\pm0.7$ &$5.6\pm0.7$  &$5.5\pm0.7$  &$5.5\pm0.6$  &$5.4\pm0.6$  &$5.5\pm0.6$   \\ \hline
\end{tabular}
\caption{\label{table:e2e_distances} Mean ($\overline{x}$) and standard deviation values ($\sigma(x)$) of the end-to-end distance $x$ of each DNA bow in both the probe-bound and probe-unbound state. Values were calculated using the measured $x$ distance values of $7.5 \times 10^4$ configurations saved over $t = \SI{1.14}{\micro\second}$ of simulation time. $x$ is defined as the distance between backbone sites on the terminal bases of the elastic arc that are covalently linked to the ssDNA target segment.}

\end{table}

\begin{table}
\renewcommand{\arraystretch}{.6} 
\centering
\begin{tabular}{|c|c|c|}
    \hline
     Order Parameter Q & Number of base pairs $n$ ($E<E_0$)\\ \hline

     $Q = -1$  & $n = 9$ \\
     $Q = 0$  & $n = 8$ \\
     $Q = 1$  & $n = 7$ \\
     $Q = 2$  & $n = 6$ \\
     $Q = 3$  & $n = 5$ \\
     $Q = 4$  & $n = 4$ \\
     $Q = 5$  & $n = 3$ \\
     $Q = 6$  & $n = 2$ \\
     $Q = 7$  & $n = 1$ \\
     $Q = 8$  & $n = 0$\\
     \hline
\end{tabular}
\caption{\label{table:unbinding_order_parameters} Order parameter definitions for FFS unbinding simulations. Each interface is defined by the value of a corresponding order parameter (e.g. separation $d$ or base pairs). Base pairs are defined as when two complementary nucleotides have a hydrogen bonding energy lower than the energy scale $E_0 = \SI{-0.6}{\kilo\calorie / \mole}$. Note that base pairs formed in misaligned duplex structures are not counted.}
\end{table}

\begin{table}
\renewcommand{\arraystretch}{.6} 
\centering
\begin{tabular}{|c|c|c|}
    \hline
     Order Parameter Q & Minimum Separation $d$/nm & Base pairs $n$ ($E<E_0)$\\ \hline
     $Q =-2$ &  $d > 3.41$ & - \\
     $Q = -1$ &  $3.41 \geq d > 1.70$ & - \\
     $Q = 0$ &   $1.70 \geq d > 0.85$ & -  \\ 
     $Q = 1$ & $d \leq 0.85$ & $n = 0$ \\
     $Q = 2$ & - & $n = 1$ \\
     $Q = 3$ & - & $n = 2$ \\
     $Q = 4$ & - & $n = 3$ \\
     $Q = 5$ & - & $n = 4$ \\
     $Q = 6$ & - & $n = 5$ \\
     $Q = 7$ & - & $n = 6$ \\
     $Q = 8$ & - & $n = 7$ \\
     $Q = 9$ & - & $n = 8$ \\     
     $Q = 10$ & - & $n = 9$ \\
     \hline
\end{tabular}
\caption{\label{table:binding_order_parameters} Order parameter definitions for FFS binding simulation. Each interface of the FFS simulation is defined using an order parameter value (e.g. separation $d$ or base pairs $n$). The symbol ‘-’ indicates that there is no constraint on the particular coordinate for the given interface. Minimum separation is defined as the minimum distance between any two complementary bases on the target and probe strands. Base pairs are defined as when two complementary nucleotides have a hydrogen bonding energy lower than the energy scale $E_0 = \SI{-0.6}{\kilo\calorie / \mole}$. Note that misaligned duplex structures are not counted as base pairs.}
\end{table}

\begin{table}
\scriptsize
  \renewcommand{\arraystretch}{.7} 
  \centering
  \begin{tabular}{|c|c|c|c|c|c|c|c|c|c|}
\hline
\multicolumn{10}{|c|}{\textbf{FFS results for probe P1 (GTAAATTCA)}} \\
\hline
\multicolumn{2}{|c|}{} & \multicolumn{8}{|c|}{\textbf{Reaction type}} \\
\hline
\multicolumn{2}{|c|}{} & \multicolumn{4}{|c|}{Unbinding} & \multicolumn{4}{|c|}{Binding} \\
\hline
\multicolumn{2}{|c|}{} & \multicolumn{8}{|c|}{\textbf{Target strand end-to-end fixed extension value}} \\
\hline
\multicolumn{2}{|c|}{} & \multicolumn{4}{|c|}{\SI{5.5}{\nano\meter}} & \multicolumn{4}{|c|}{\SI{5.1}{\nano\meter}} \\
\hline
\textbf{Trial} & \textbf{Goal Interface} & \multicolumn{8}{|c|}{\textbf{Number of forward crossings}} \\
\hline
1 & & \multicolumn{4}{|c|}{30017} & \multicolumn{4}{|c|}{30002} \\
2 & $\lambda_{0}$ & \multicolumn{4}{|c|}{60013} & \multicolumn{4}{|c|}{60001}\\
3 & & \multicolumn{4}{|c|}{60015} & \multicolumn{4}{|c|}{60000}\\
\hline
\textbf{Trial} & \textbf{Goal Interface} & \multicolumn{8}{|c|}{\textbf{Initial Flux ($\mathrm{ns}^{-1}$)}} \\
\hline
1 & & \multicolumn{4}{|c|}{57.3} & \multicolumn{4}{|c|}{0.515} \\
2 & $\lambda_{0}$ & \multicolumn{4}{|c|}{56.6} & \multicolumn{4}{|c|}{0.518}\\
3 & & \multicolumn{4}{|c|}{57.4}& \multicolumn{4}{|c|}{0.520}\\
\hline
\textbf{Trial} & \textbf{Goal Interface} & \textbf{Successes} & \textbf{Prune} & \textbf{Attempts} & \textbf{Prob.} & \textbf{Successes} & \textbf{Prune}& \textbf{Attempts} & \textbf{Prob.}\\

 & & & \textbf{Successes} & & & & \textbf{Successes}& & \\
\hline
1 & & 30001 & - & 857297 & 0.035 & 30001 & - & 1309017 & 0.023 \\
2 & $\lambda_{1}$ & 30000 & - & 861360 & 0.035 & 30000 & - & 1308931 & 0.023 \\
3 & & 30001 & - & 850404 & 0.035 & 30000 & - & 1302569 & 0.023 \\
\hline
1 & & 30002 & 15223 & 1299683 & 0.058 & 15585 & 859 & 6864816 & 0.003 \\
2 & $\lambda_{2}$ & 30001 & 15048 & 1284311 & 0.059 & 12230 & 928 & 5173098 & 0.003 \\
3 & & 30000 & 15167 & 1276431 & 0.059 & 13688 & 924 & 5733836 & 0.003 \\
\hline
1 & & 30004 & 17169 & 1543954 & 0.053 & 30001 & 93 & 174849 & 0.173 \\
2 & $\lambda_{3}$ & 30003 & 17306 & 1539043 & 0.053 & 30001 & 113 & 168188 & 0.18 \\
3 & & 30001 & 17395 & 1491069 & 0.055 & 30000 & 112 & 165547 & 0.183 \\
\hline
1 & & 30001 & 17373 & 1837336 & 0.045 & 30000 & 17857 & 135201 & 0.618 \\
2 & $\lambda_{4}$ & 30002 & 17244 & 1799833 & 0.045 & 30001 & 17697 & 130722 & 0.636 \\
3 & & 30003 & 17451 & 1728580 & 0.048 & 30002 & 17233 & 125808 & 0.649 \\
\hline
1 & & 30001 & 17501 & 1930017 & 0.043 & 30012 & 17036 & 194725 & 0.942 \\
2 & $\lambda_{5}$ & 30001 & 17707 & 1871701 & 0.044 & 30005 & 16393 & 185757 & 0.956 \\
3 & & 30000 & 18032 & 1829295 & 0.046 & 30012 & 15795 & 180810 & 0.952 \\
\hline
1 & & 30000 & 19898 & 1693959 & 0.053 & 30018 & 15787 & 173298 & 0.993 \\
2 & $\lambda_{6}$ & 30000 & 20882 & 1489686 & 0.062 & 30006 & 16457 & 179119 & 0.994 \\
3 & & 30001 & 20541 & 1565933 & 0.059 & 30014 & 16098 & 176816 & 0.989 \\
\hline
1 & & 10011 & 6924 & 554063 & 0.056 & 30021 & 22227 & 96261 & 1.005 \\
2 & $\lambda_{7}$ & 10004 & 7206 & 499387 & 0.063 & 30010 & 22622 & 98389 & 0.995 \\
3 & & 6796 & 4788 & 370763 & 0.057 & 30022 & 22178 & 97357 & 0.992 \\
\hline
1 & & 10001 & 7386 & 832578 & 0.039 & 30022 & 20175 & 90038 & 1.006 \\
2 & $\lambda_{8}$ & 9114 & 6605 & 811829 & 0.036 & 30011 & 20179 & 90944 & 0.996 \\
3 & & 5000 & 3596 & 382574 & 0.041 & 30021 & 20319 & 91490 & 0.994 \\
\hline
1 & & 10007 & 5000 & 256882 & 0.097 & 30022 & 18749 & 86244 & 1.0 \\
2 & $\lambda_{9}$ & 10001 & 5030 & 255935 & 0.098 & 30011 & 18767 & 86104 & 1.002 \\
3 & & 5002 & 2576 & 127468 & 0.1 & 30022 & 18803 & 86856 & 0.995 \\
\hline
1 & & - & - & - & - & 30022 & 18571 & 85373 & 1.004 \\
2 & $\lambda_{10}$ & - & - & - & - & 30023 & 18355 & 85408 & 0.996 \\
3 & & - & - & - & - & 30022 & 18628 & 85253 & 1.008 \\
\hline
\hline
\multicolumn{2}{|c|}{} & \multicolumn{4}{|c|}{} & \multicolumn{4}{|c|}{} \\
\multicolumn{2}{|c|}{\textbf{$k_{AB}$}} & \multicolumn{4}{|c|}{\textbf{$\SI{9.9}{\per \min}$}} & \multicolumn{4}{|c|}{\textbf{$\SI{2.3}{\per \second \per \micro\molar}$}} \\
\multicolumn{2}{|c|}{} & \multicolumn{4}{|c|}{} & \multicolumn{4}{|c|}{} \\
\hline
\end{tabular}\caption{Forward flux simulation results for binding and unbinding reactions. The ``Successes" column specifies the number of configurations that successfully cross the goal interface $\lambda_{Q}$ coming from $\lambda_{Q-1}$; ``Prune Successes" specifies the number of configurations that survive pruning upon traveling backward to $\lambda_{Q-2}$ from $\lambda_{Q-1}$, and afterward cross $\lambda_{Q}$. Trajectories that revert to $\lambda_{Q-2}$ were pruned with a $p=0.75$ probability. ``Attempts" specifies the total number of trajectories started at $\lambda_{Q-1}$. The probability (``Prob") of crossing $\lambda_0$ was calculated using Equation $\ref{eq:melt_prob}$, while the probability of crossing the remaining interfaces was calculated using Equation $\ref{eq:prune_melt_prob}$. The final rate $k_{AB}$ was then calculated using Equation \ref{eq:ffs}. \label{table:ffs_results}}
\end{table}

\clearpage
\newpage
\end{document}